# MedBuild AI: An Agent-Based Hybrid Intelligence Framework for Reshaping Agency in Healthcare Infrastructure Planning through Generative Design for Medical Architecture

A PREPRINT

Yiming Zhang[1]; Yuejia Xu[1]; Ziyao Wang[1]; Xin Yan[1]*; Xiaosai Hao[1]

[1] School of Architecture and Urban Planning, Beijing University of Civil Engineering and Architecture (BUCEA), Beijing 100044, China

* Corresponding author: yanxin@bucea.edu.cn

Yiming Zhang: https://orcid.org/0009-0008-3488-4656

Yuejia Xu: https://orcid.org/0000-0002-4039-1734

Xin Yan: https://orcid.org/0000-0002-5033-3597

### Abstract

Globally, disparities in healthcare infrastructure remain stark, leaving countless communities without even basic services. Traditional infrastructure planning is often slow and inaccessible, and although many architects are actively delivering humanitarian and aid-driven hospital projects worldwide, these vital efforts still fall far short of the sheer scale and urgency of demand. This paper introduces MedBuild AI, a hybrid-intelligence framework that blends large language models (LLMs) with deterministic expert systems to rebalance the early design and conceptual planning stages. As a web-based platform, it enables any region with satellite internet access to obtain guidance on modular, low-tech, low-cost medical building designs. The system operates through three agents: the first gathers local health intelligence via conversational interaction; the second translates this input into an architectural functional program through rule-based computation; and the third generates layouts and 3D models. By embedding computational negotiation into the design process, MedBuild AI fosters a reciprocal, inclusive, and equitable approach to healthcare planning—empowering communities and redefining agency in global healthcare architecture.

## 1. Introduction

Achieving health equity, as advocated by United Nations Sustainable Development Goal 3, is a formidable challenge in today's world. According to the latest report from the World Health Organization and the World Bank, global progress on Universal Health Coverage (UHC) has slowed significantly in recent years: the UHC Service Coverage Index only increased from 45 to 68 between 2000 and 2021, and has largely stagnated between 2019 and 2021 [1]. At this rate, it is estimated that approximately 4.5 billion people worldwide were not fully covered by essential health services in 2021 [1]. Concurrently, financial protection has continued to deteriorate, with about 2 billion people facing financial hardship in 2019 due to high out-of-pocket health expenditures [1]. To visually represent this service coverage gap, we have mapped the global UHC Service Coverage Index for 2021 (Figure 1), based on data from the WHO Global Health Observatory [2] and following the index construction methodology outlined in the global monitoring report [1]. This combination of "coverage stagnation" and "worsening financial hardship" highlights a dual dilemma in resource-scarce regions: communities not only lack medical facilities but also the specialized expertise to plan these complex buildings, resulting in both a "knowledge gap" and a "resource gap."

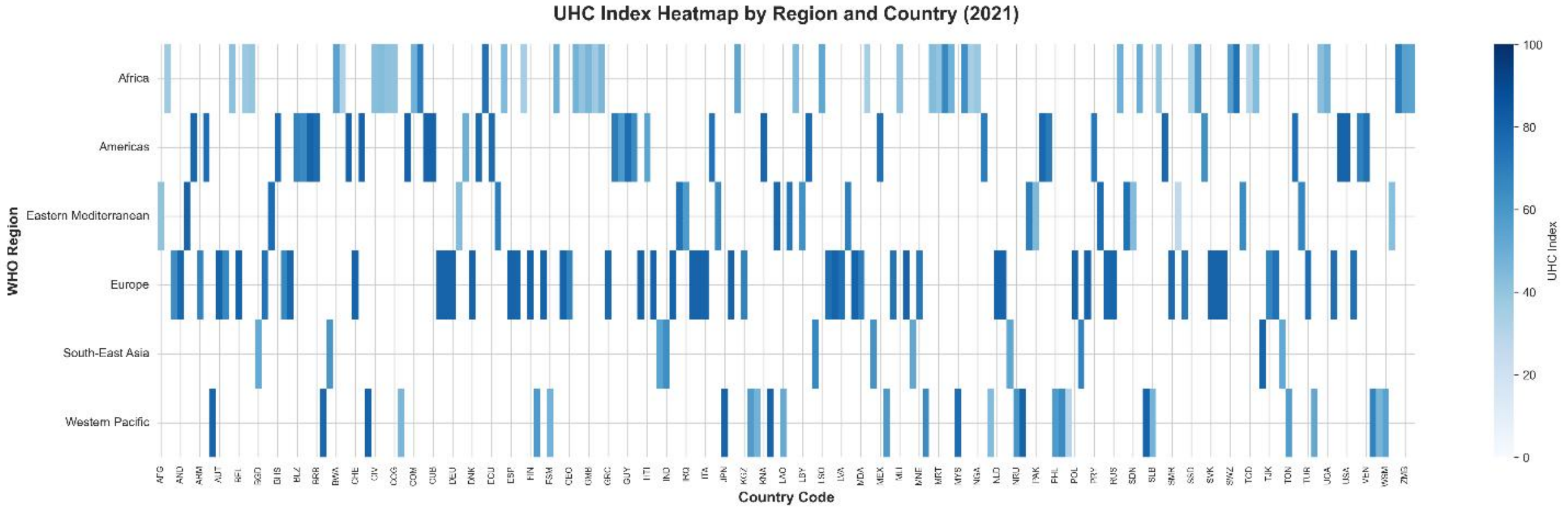

*Figure 1: UHC Service Coverage Index (SCI) by country, 2021*

The African continent is an extreme microcosm of this predicament. Africa is home to nearly 17% of the world's population but bears 24% of the global disease burden, while its health expenditure is less than 1% of the global total, and its health workforce constitutes only 3% of the world's total [3], [4]. Studies show that healthcare resources are highly concentrated in urban areas, with rural regions having only about 23% of doctors and 38% of nurses [3]. This severe resource mismatch leads to a significant "spatial imbalance," where vast rural areas become "healthcare deserts" lacking basic services. This spatial imbalance can be quantified and visualized using advanced geographic information tools. Figure 2 is an example captured using the Open Healthcare Access Map application [5], which integrates road network data from OpenStreetMap [6], population grid data from WorldPop [7], and administrative boundaries from geoBoundaries [8] to intuitively present the accessibility situation, using the Democratic Republic of Congo as an example. This situation not only systematically infringes upon the right to health equity for vulnerable populations but also poses a severe challenge to the healthcare system's response capacity in emergencies such as pandemics [9].

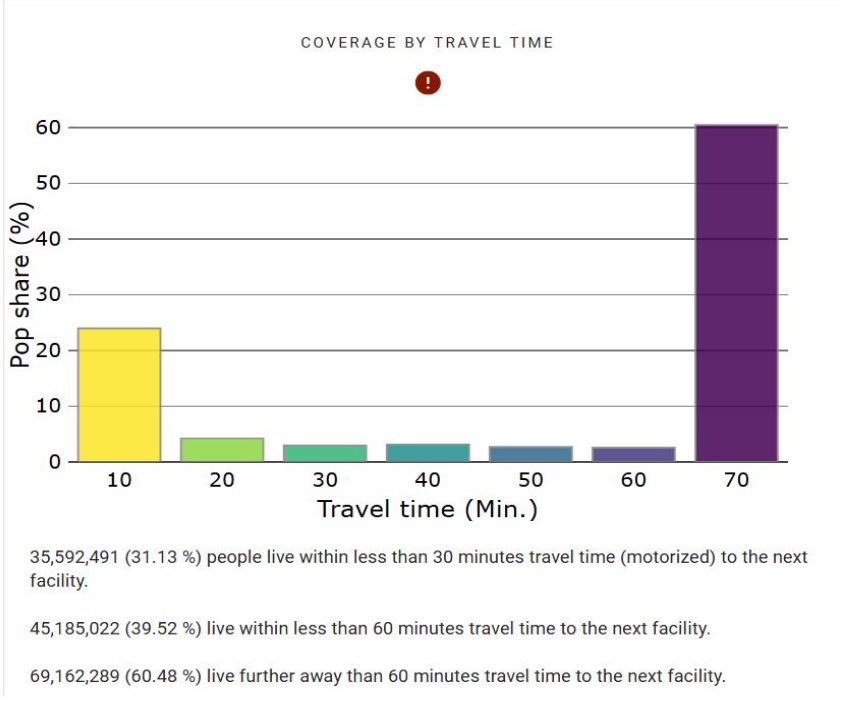

Coverage by travel time to nearest healthcare (Democratic Republic of the Congo, DRC)
© openrouteservice.org by HeiGIT (CC BY 4.0) | © OpenStreetMap contributors (ODbL 1.0) | WorldPop (CC BY 4.0) | geoBoundaries (CC BY 4.0)
Screenshot by author using the Open Healthcare Access Map; accessed 2025-10-01

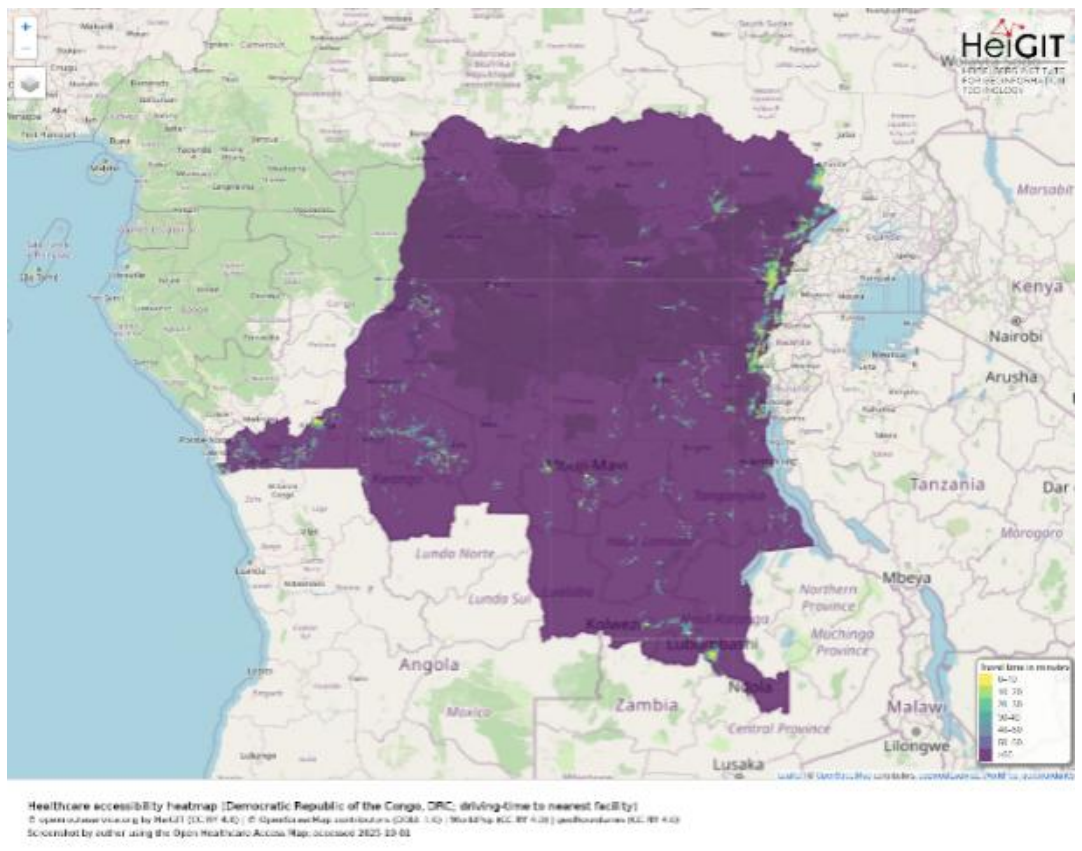

*Figure 2: Spatial imbalance in healthcare service accessibility in the Democratic Republic of Congo (motorized travel, binned 0–10…70+ minutes) Screenshot generated by the author using the Open Healthcare Access Map application [5]*

Against this backdrop, traditional healthcare facility planning paradigms often fall into a so-called "impossible triangle" due to complex local contextual challenges [10]—an inherent conflict between design cycle, construction cost, and contextual fit. This is not just a technical problem but reflects a systemic "interaction failure": purely relying on mathematical optimization or traditional expert-driven linear processes cannot effectively address the complex reality composed of hard-to-quantify cultural, social, and economic factors [11], [12].

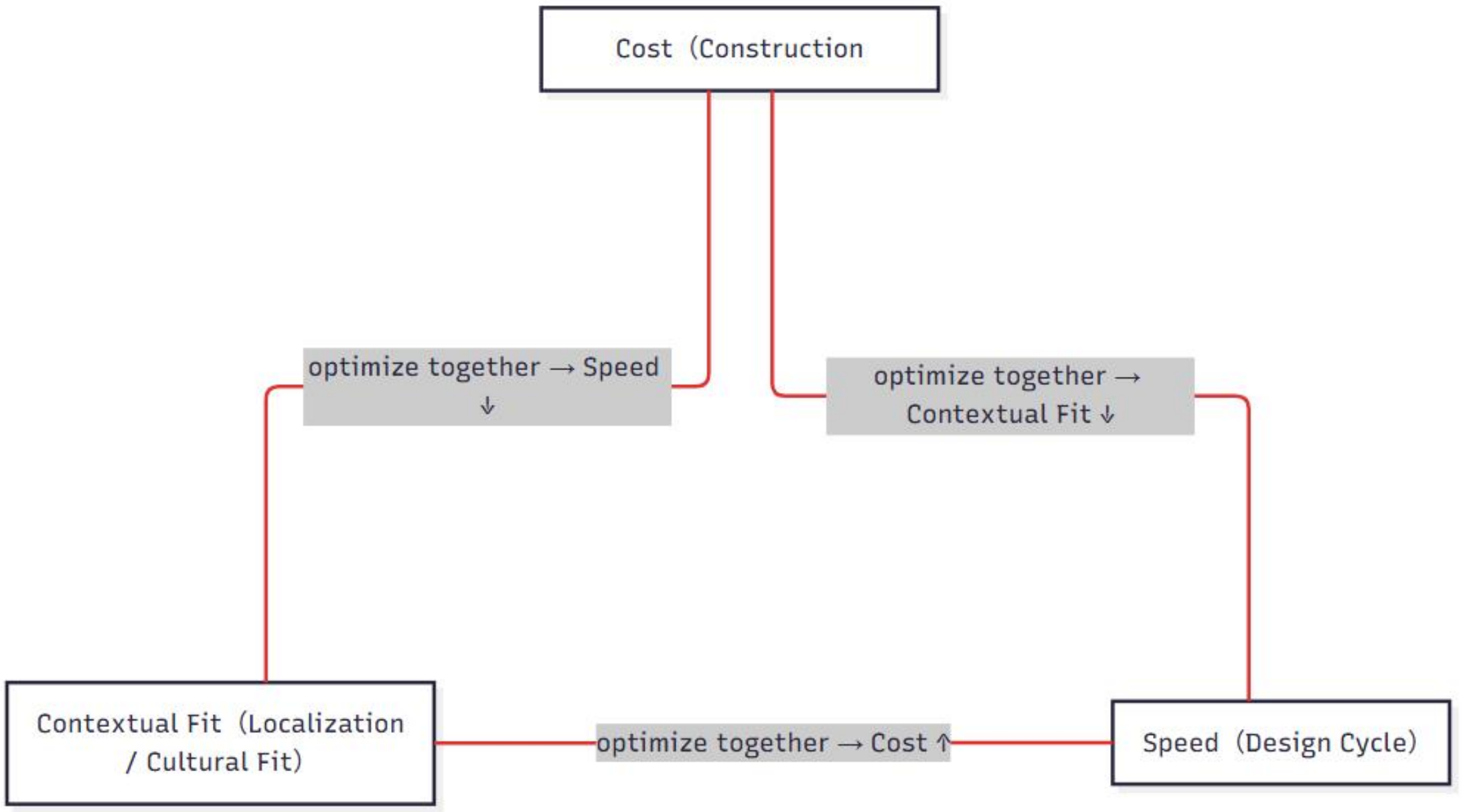

*Figure 3: The Impossible Triangle*

This interaction failure disempowers local communities and undermines project sustainability, manifesting as: Information Gap: Qualitative local knowledge is systematically marginalized, and planning tends to apply generic international standards, leading to facilities that are disconnected from the community's actual needs. Communication Barriers: Obstacles arising from multilingualism and social customs not only cause the loss of critical information but also strip local stakeholders of their agency [13]. Structural Inefficiency: The lengthy approval processes in aid projects often compress the pre-design phase, sacrificing the in-depth, participatory fieldwork necessary for true localization [14]. Constructability Dilemma: Design solutions often rely on scarce imported materials and complex technologies, overlooking modular, low-tech construction systems that are better suited to the local context.

Recent developments in Artificial Intelligence (AI) offer a historic opportunity to address this. However, a responsible path is not to build an omnipotent, singular AI, but to create a hybrid intelligence framework that promotes critical reflection on human-machine roles [15]. The potential of AI lies in its ability to act as a "translator" and "enabler" of knowledge, transforming complex professional expertise into tools that local communities can use. Research emphasizes that the participation of local communities in the data creation and model training of AI systems is crucial for ensuring algorithmic fairness and accountability [16].

In this context, we introduce MedBuild AI, which is not just a tool for generating design solutions, but a medium for design negotiation. Based on this framework, we have developed MedBuild 1.0, a web-accessible design platform ( available at https://intro-dot-medbuildai.uc.r.appspot.com/ ). The platform is structured as a three-agent system that establishes an end-to-end workflow from qualitative dialogue to interactive spatial synthesis. Its core contribution is to reshape agency and promote computational reciprocity by creating a transparent, shared platform where indigenous knowledge becomes the cornerstone of the design process, and the complex trade-offs between cost, culture, and clinical function are made legible to all stakeholders.

This paper will first review the history of computational design in the field of medical architecture; then, it will detail the three-agent methodology of MedBuild AI; subsequently, it will demonstrate the framework's adaptability through a series of case studies; finally, it will discuss the implications, limitations, and future directions of this approach, aiming to provide a new paradigm for a more equitable and contextually responsive design practice.

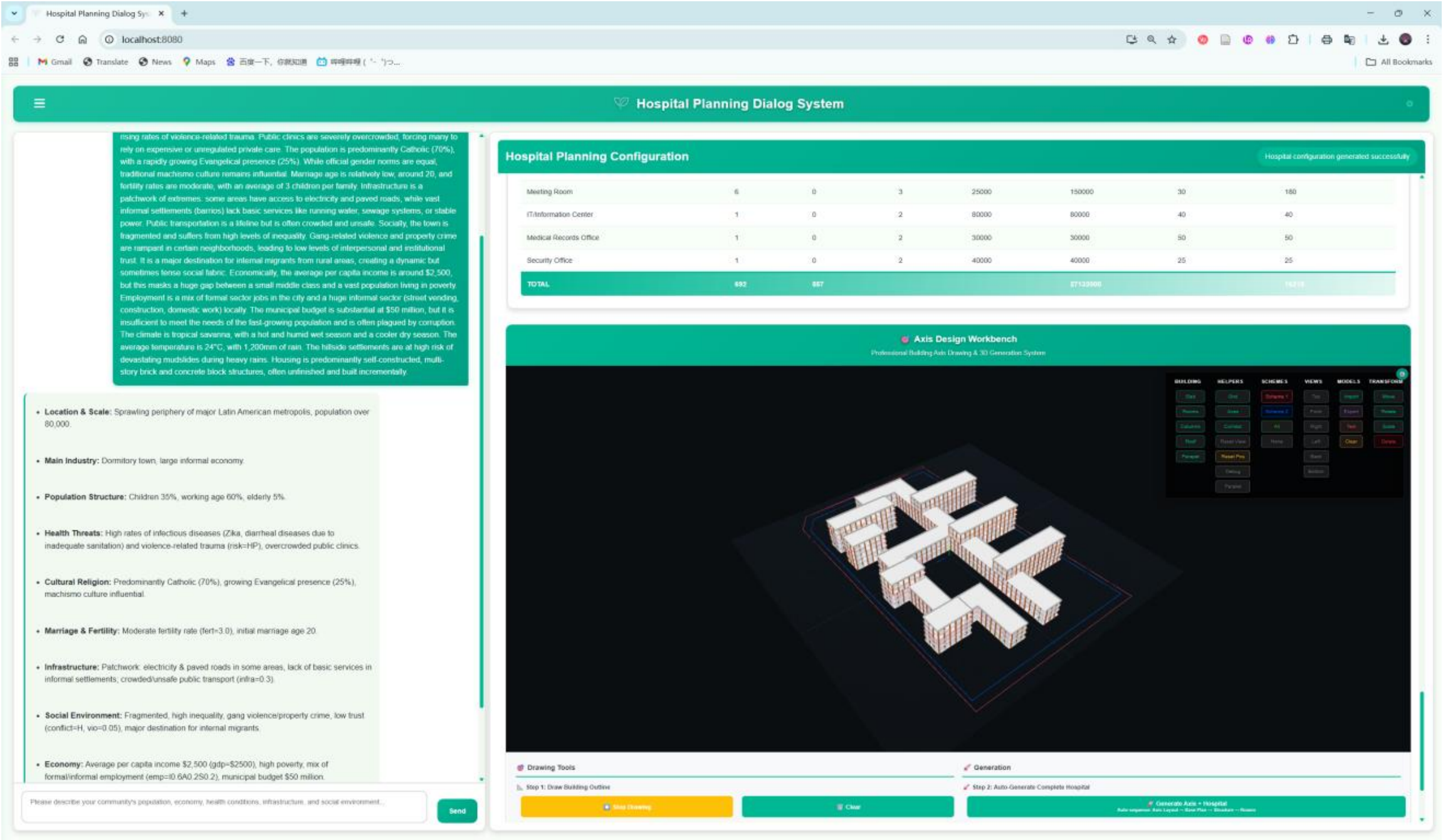

*Figure 4: system UI*

## 2. Related Work

The architectural design of healthcare facilities is an extremely complex systems engineering problem, requiring a delicate balance between functional efficiency, operational costs, patient experience, and staff well-being [16]. To address this challenge, computational design methods have continuously evolved over the past few decades [17], providing designers with a rich toolkit ranging from local optimization to systemic generation. However, these research outcomes have largely manifested as "point" solutions targeting specific segments of the design process, failing to form a coherent, integrated workflow. It is against this historical backdrop that this study proposes an integrated hybrid intelligence framework, marking a critical turning point from an isolated toolkit to a collaborative agent-based system.

### 2.1. Early Generative Paradigms: From Mathematical Optimization to Rule-Based Systems

The computational exploration of healthcare facility layout planning traces its roots to the Facility Layout Planning (FLP) problem in industrial engineering, with the core objective of minimizing the cost and time of transporting personnel and materials by optimizing the spatial arrangement of departments. One of the key early methodologies was Systematic Layout Planning (SLP), proposed by Muther in 1973 [18], a procedural, qualitative relationship diagram-based process that remains influential today and laid the logical foundation for subsequent computational models. This approach of decomposing a design problem into function, behavior, and structure also resonates with the classic Function-Behavior-Structure (FBS) framework in design theory [19].

With the advent of computational technology, researchers formalized this problem into mathematical optimization models, ushering in an era of automated layout generation and optimization. The Quadratic Assignment Problem (QAP), proposed by Koopmans and Beckmann in 1957 [20], became a classic model in this field. Elshafei first applied it in 1977 to optimize the layout of hospital clinics to reduce patient travel distance, pioneering the use of mathematical optimization in medical architecture design [21]. To address the limitations of QAP in handling real-world problems, recent research has continued to advance. For instance, Cubukcuoglu et al. (2021) combined QAP with geodesic distances for hospital renovation projects [12], and Cetintas (2024) further explored QAP-based generative modeling methods [22], demonstrating the enduring vitality of this classic model.

To overcome the QAP's limitation of assuming equal departmental areas, Mixed-Integer Programming (MIP) was introduced, allowing for the handling of departments of different sizes and shapes and providing more flexible layout solutions. Cubukcuoglu et al. (2022) constructed a computational workflow that included stacking, zoning, and circulation planning, showcasing the powerful capabilities of MIP in handling complex new hospital projects [11]. However, both QAP and MIP are NP-hard problems with extremely high computational complexity. Consequently, subsequent research turned to heuristic and meta-heuristic algorithms, such as the Genetic Algorithm (GA), which can find near-optimal solutions in a reasonable amount of time. For example, Chen et al. (2024) applied a genetic algorithm combined with patient flow analysis to optimize hospital layouts [23]. Furthermore, in the face of multiple conflicting objectives common in design (such as cost, efficiency, and daylighting), multi-objective/non-linear optimization emerged. Researchers have used evolutionary parametric tools to optimize spatial adjacencies [24] or to solve the unequal-area healthcare facility layout problem while considering performance indicators like solar access [25]. The "Space Plan Generator" developed by Das et al. (published at ACADIA 2016) is a typical representative of this approach [26]; it can rapidly generate and evaluate multiple floor plan options, providing a basis for decision-making. These computational methods have also been applied to the design of highly specialized medical spaces such as robotic surgery rooms [27].

Parallel to the development of optimization algorithms, another path was based on rule-based generative systems, which attempt to directly encode the intrinsic logic of design. Shape Grammar, proposed by Stiny and Gips in the 1970s [28], uses a set of predefined shape rules to analyze and generate architectural designs with specific styles. Although its direct application in medical architecture is limited, its idea of encoding design logic into computable "digital genes" provided a crucial conceptual inspiration for all subsequent generative models and echoes broader theories of algorithmic architecture [29]. These early generative paradigms, whether based on mathematical optimization or manual rules, laid the foundation for automated design exploration, but they were essentially "closed systems," lacking the ability to learn and adapt from real-world cases.

### 2.2. The Data-Driven Turn: AI Generation and Human-Centric Evaluation

In recent years, with breakthroughs in deep learning, the research paradigm has been shifting from "rule- and optimization-based" to "data- and learning-based" "open systems" [30]. AI generative methods, represented by Generative Adversarial Networks (GANs), have brought new possibilities to architectural layout design. They no longer rely on preset rules but learn latent patterns and styles directly from large amounts of real design data [31], [32], and have been used to generate architectural images with a specific sense of place [33]. In the field of medical architecture, Zhao et al. (2021) pioneered the application of GANs (including pix2pix and CycleGAN) to generate layout plans for hospital emergency departments [34], demonstrating the potential of AI to learn from existing design cases and create new, functionally sound layouts. To better capture the complex topological relationships between rooms, Graph Neural Networks (GNNs) have also been introduced for layout generation, such as the Graph2Plan framework proposed by Hu et al. (2020) [35]. In addition, emerging techniques like self-supervised learning have been used to automatically generate building archetypes for subsequent performance analysis (e.g., energy consumption) [36]. These AI generation methods have greatly enhanced the breadth and speed of design exploration, but their "black-box" nature, reliance on large-scale, high-quality datasets, and challenges in handling hard constraints have limited their direct application [37]. More importantly, data-driven methods themselves may inherit and amplify biases present in existing data, thus requiring a critical perspective to examine their ethical boundaries [38].

While AI models focus on generating layout schemes, another important line of research concentrates on evaluating the human impact of these schemes. The Space Syntax theory, founded by Hillier and Hanson [39], quantifies and analyzes topological properties of space such as connectivity, integration, and intelligibility by abstracting architectural space into a network graph. In the healthcare domain, Space Syntax is widely used to evaluate and improve hospital wayfinding efficiency, nurses' movement patterns, and staff communication and interaction, and has led to new evaluation metrics to measure the impact of design on the quality of care [40]. It is not a generative method but provides a crucial human-centric evaluation dimension for generative design. To better connect abstract spatial analysis with concrete design entities, Jia et al. (2025) proposed the semantic Configuration Model (HCM) [41] as a more advanced way of representing hospital spatial information, providing a solid model foundation for design support systems to ensure that generated schemes are not only functionally rational but also easy to use and experience.

### 2.3. Research Gap and Paradigm Shift

In summary, computational design in medical architecture has evolved from optimization- and rule-based generation to data- and learning-based generation, supplemented by space topology-based human-centric evaluation. Each class of methods has made significant contributions to solving specific design challenges. However, a glaring gap remains: these powerful toolchains are still mutually disconnected. In the design process, a significant chasm exists between quantitative optimization (e.g., QAP/MIP), data-driven pattern generation (e.g., GANs), and qualitative, negotiated local needs (e.g., culture, resource constraints). The entire design process still relies on human designers to manually "translate" and make decisions between different tools and knowledge domains, which is inefficient and prone to creating information silos.

We are at a critical turning point. On one hand, the theories and tools for the various computational methods mentioned above have become increasingly mature. On the other hand, a new generation of AI technologies, represented by Large Language Models (LLMs), has demonstrated unprecedented capabilities in natural language understanding, knowledge integration, and logical reasoning, making them ideal "conductors" for connecting and coordinating different professional tools. This fusion of the pattern recognition capabilities of neural networks with the logical reasoning of symbolic systems, known as Neuro-Symbolic AI, is considered the "third wave of AI" following deep learning [42], and AI-based hospital design processes based on this strategy are already on the horizon for researchers [43]. Although integrating these capabilities into a coherent agent-based system is still a frontier exploration in medical architecture, this paradigm shift is already beginning to emerge in the broader fields of architectural design, 3D modeling, and computer-aided design (CAD). Recent studies indicate that Multi-Agent architectures are becoming central to achieving end-to-end automation and human-machine collaboration, with reasoning models for exploratory computational design appearing [44].

For example, researchers have proposed various LLM-driven agent frameworks that can directly translate a designer's sketch or natural language description into complex 3D models [45], [46], [47], [48], or be used to generate layouts for other complex building types like airport terminals [50]. The multiple agents within these systems work collaboratively, with some responsible for visual understanding, others for logical reasoning and code generation, and still others for self-reflection and correction [45], [51]. Among these, some frameworks like "Building-Agent" combine the planning ability of LLMs with the powerful spatial generation capabilities of Graph Neural Networks (GNNs) [52], [53]; others like "BIMgent" explore having AI agents directly operate the user interfaces of existing BIM software [54]; and still other research utilizes LLMs and knowledge graphs to enhance the interpretability of IFC models [55], aiming to bridge the entire process from concept to professional model. Furthermore, LLMs are also being used as a "mediator" or "co-pilot" between designers and complex computational tools [56], [57], enabling parametric modeling in mixed reality (MR) environments through conversational interfaces [58], or combining neural reasoning with traditional mathematical solvers to complement AI creativity with computational rigor [58]. The remarkable logical reasoning ability demonstrated by AI in solving Olympiad-level geometry problems [59] also foretells its future potential in handling complex design constraints such as building codes [60]. These advancements in non-medical architecture clearly signal the huge potential of agent-based systems as the next generation of design tools and provide a strong reference for this study. The fact that top academic conferences like CAADRIA 2025 have dedicated sessions on "Agent-based Generative Design" signifies that this direction has become a frontier hotspot in computational design.

The MedBuild AI framework proposed in this paper is a response to this research gap. By constructing a Hybrid Intelligence system composed of a conversational AI, a multi-agent LLM pipeline, an expert system, and a generative model, it attempts for the first time to proceduralize, automate, and intelligentize the entire pre-design and conceptual phases of medical architecture. Its core contribution lies not only in the technical "linking" but also in recalibrating the relationship between global knowledge and local reality through the reciprocal relationship between the LLM and the expert system, as well as the negotiated dialogue of human-machine collaboration. This profound human-computer interaction model [61] aims to achieve a more inclusive design process [62] and critically reflect on the role of "participation" in design [63], thereby reshaping agency in the design process and providing more adaptive and humanistic medical building solutions for resource-scarce regions.

## 3. Methodology: The MedBuild AI Three-Agent Framework

At its core, MedBuild AI is a modular, service-oriented web application architected as a multi-agent system. This architecture facilitates a "dialogue-compile-calculate-generate-synthesize" workflow, strategically decomposing the complex design problem into specialized tasks handled by three distinct, sequential agents. This structure ensures a clear separation of concerns, maintainability, and the robust integration of different AI paradigms. The central nervous system of the platform is the Dimensional Qualitative Local-health-intelligence (DQL), a standardized JSON-like string that acts as the project's "digital gene," ensuring data integrity and interoperability between all agents as information is progressively refined.

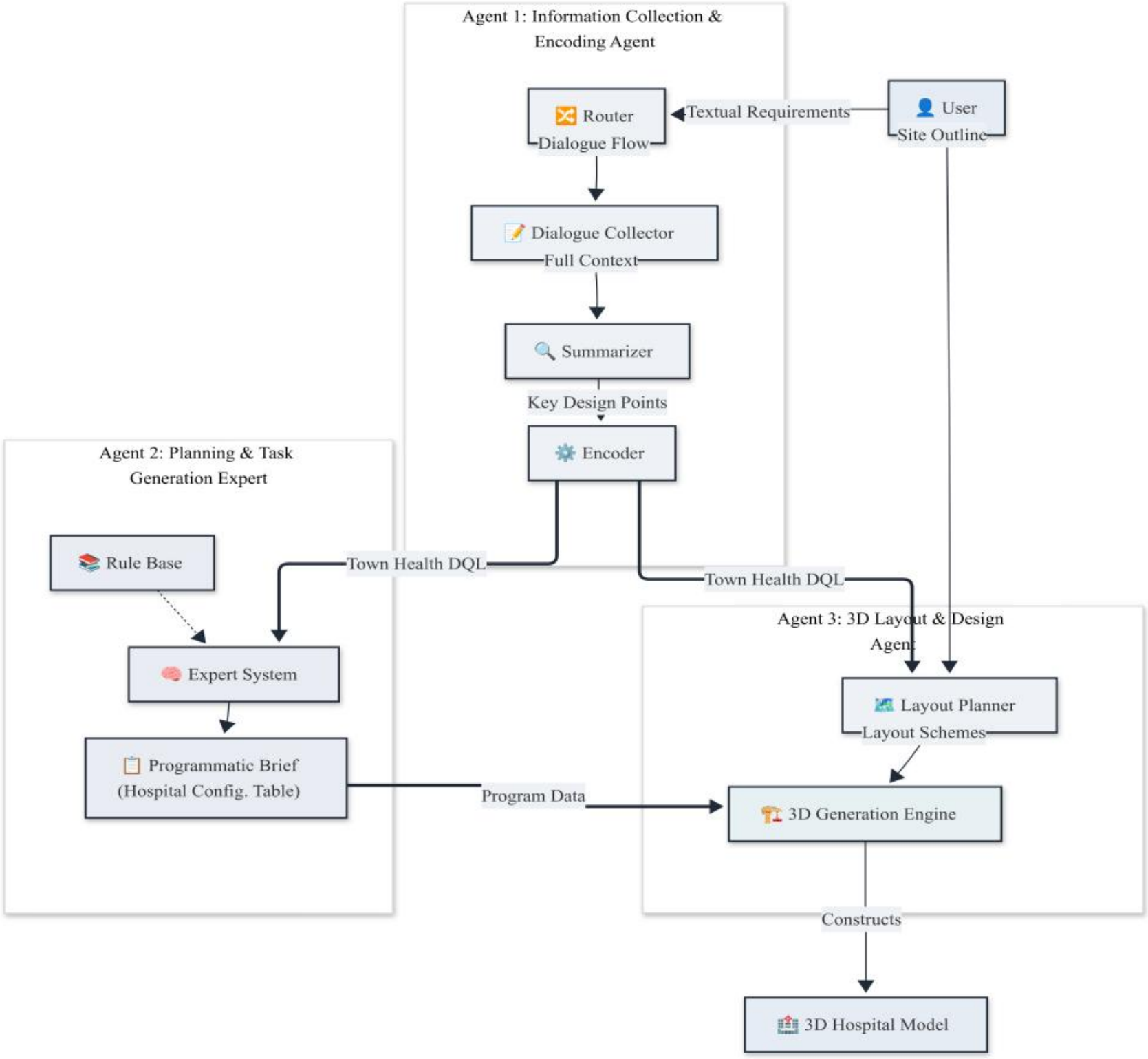

*Figure 5: MedBuild AI System Architecture Flowchart*

### 3.1 Agent 1: Dialogue & Information Encoding Agent

This agent serves as the critical human-computer interface, with its core task being to perform a complex data dimensionality reduction process. It is responsible for transforming unstructured, qualitative local health intelligence (high-dimensional data) into a structured, machine-readable DQL (low-dimensional data). It functions as a "semantic compiler" through a four-stage internal pipeline, compressing and encoding human narrative wisdom (high-dimensional, unstructured information) into precise instructions that a machine can understand. This transformation process is crucial for ensuring that local knowledge, while retaining its core value, can be efficiently utilized by the computational system.

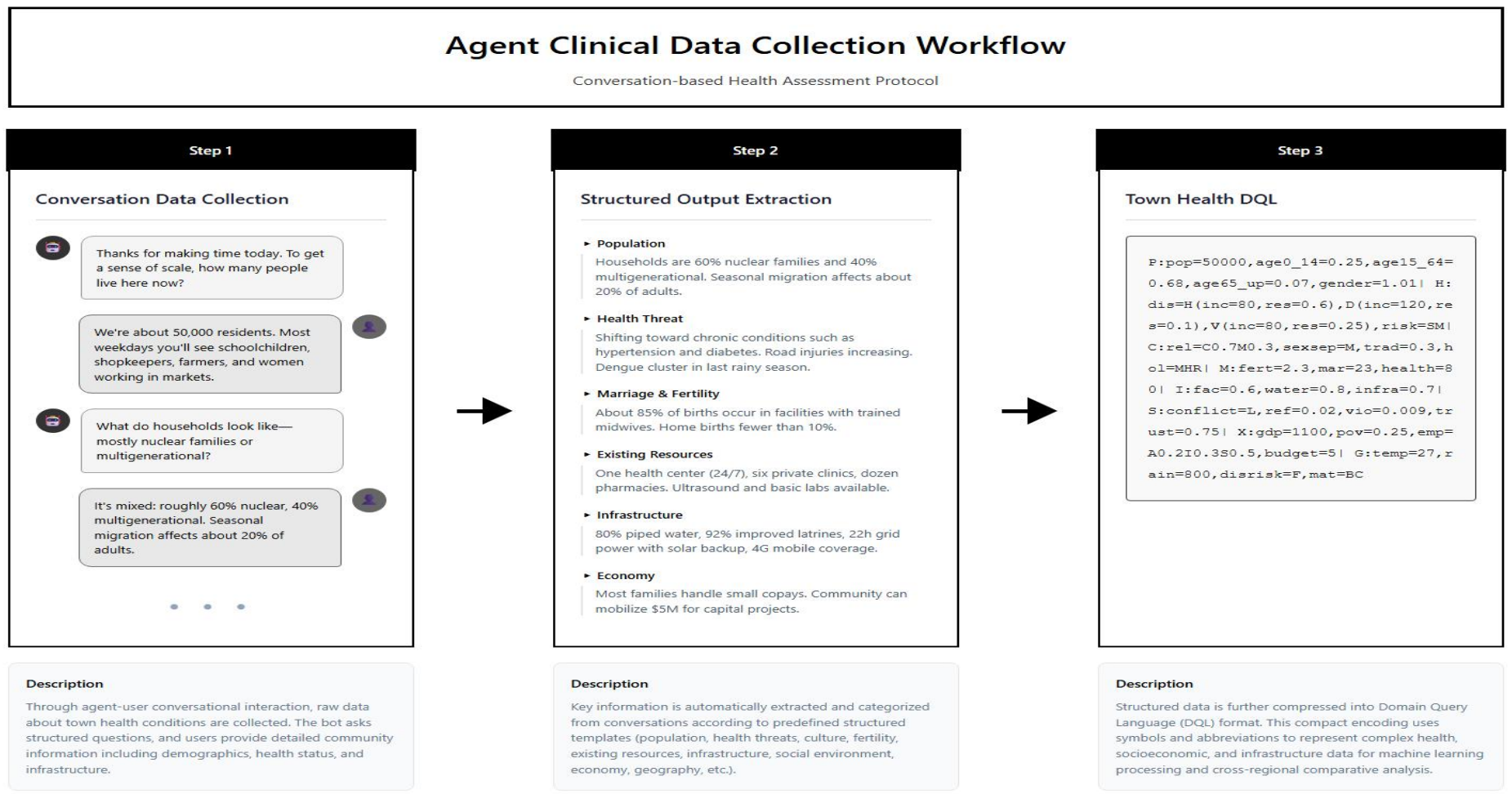

*Figure 6: Agent 1 Information Processing Pipeline*

*This agent serves as the critical human-computer interface, with its core task being to perform a complex data dimensionality reduction process. It functions as a "semantic compiler," responsible for transforming unstructured, qualitative local health intelligence (high-dimensional data) into a structured, machine-readable DQL (low-dimensional data). The workflow begins with heuristic data elicitation, where a conversational agent, MedBuild-GPT, engages the user in a Socratic dialogue. Guided by the PHCMEISXG framework, the agent poses open-ended, context-aware questions to holistically gather high-dimensional, localized information. This process is designed to be empathetic and avoids jargon, ensuring the conversation is accessible to non-expert participants and can effectively capture the nuances of human narrative wisdom.*

*Following the dialogue, the raw conversation transcript undergoes hierarchical information transformation within a two-stage compilation pipeline designed for dimensionality reduction. First, an "Integrator" LLM agent reads the entire conversation and summarizes it into a structured list of bullet points organized by the PHCMEISXG dimensions. This initial stage refines the lengthy narrative into key factual points, achieving data reduction by eliminating redundancy while preserving core semantics. Subsequently, a second, highly constrained "Encoder" LLM agent receives this summary. Using few-shot learning and examples of multiple correct input-output pairs, it further "encodes" these points into a dense, parameterized DQL string. As shown in the example below, this process is the critical step of data dimensionality reduction:*

*P:pop=50000,age0_14=0.25,age15_64=0.68,age65_up=0.07,gender=1.01|*
*H:dis=H(inc=80,res=0.6),D(inc=120,res=0.1),V(inc=80,res=0.25),risk=SM|*
*C:rel=C0.7M0.3,sexsep=M,trad=0.3,hol=MHR|M:fert=2.3,mar=23,health=80|*
*I:fac=0.6,water=0.8,infra=0.7|*
*S:conflict=L,ref=0.02,vio=0.009,trust=0.75|X:gdp=1100,pov=0.25,emp=A0.2I0.3S0.5,budget=5|*
*G:temp=27,rain=800,disrisk=F,mat=BC.*

*It compresses structured text points into an extremely compact, machine-processable format that retains all key planning variables.*

*Finally, the DQL string, now low-dimensional yet information-dense, is passed on for deterministic program generation by Agent 2, a Python-based expert system. This rule-based, deterministic module receives the encoded input and executes a series of rigorous calculations, which are detailed in Section 4, to generate the final architectural program. This critical architectural decision to place non-negotiable calculations within a deterministic system is intentional. It safeguards the core planning logic from the potential "hallucinations" of generative AI, thereby ensuring the safety, reliability, and verifiability of the outcomes.*

**Table 1: The PHCMEISXG Framework for the Dimensional Qualitative Local-health-intelligence (DQL)**

| Dimension | Name & Description | Key Parameters (Examples) |
|---|---|---|
| **P** | **Population:** Demographics and social statistics of the target community. | `pop` (total population), `age0_14`, `age15_64`, `age65_up` (age distribution), `growth_rate`, `gender` (gender ratio). |
| **H** | **Health Threat:** Local disease profiles, health risks, and epidemiological patterns. | `dis` (disease profile with incidence `inc` and resource factor `res`), `risk` (special risks, e.g., high pollution `HP`). |
| **C** | **Culture & Religion:** Dominant cultural norms, religious beliefs, and social customs. | `rel` (religious composition, e.g., M0.8C0.2 for 80% Muslim, 20% Christian), `sexsep` (gender separation needs), `trad` (reliance on traditional medicine). |
| **M** | **Marriage & Fertility:** Patterns related to maternal and child health. | `fert` (fertility rate), `mar` (average marriage age), `health` (maternal & child health index). |
| **E** | **Existing Resources:** Current state of local healthcare facilities and personnel. | `total_beds` (existing bed count), `quality_factor` (quality coefficient of existing facilities), `or_rooms` (existing operating rooms). |
| **I** | **Infrastructure:** Availability and quality of public utilities and transportation. | `fac` (healthcare facility coverage rate), `water` (water security index), `infra` (transport accessibility). |
| **S** | **Social Environment & Security:** Social cohesion, conflict levels, and safety conditions. | `conflict` (level of social conflict), `ref` (refugee proportion), `vio` (violence rate), `trust` (social trust level). |
| **X** | **Economy:** The community's economic capacity and project budget. | `gdp` (GDP per capita), `pov` (poverty rate), `emp` (employment structure), `budget` (total construction budget in millions USD). |
| **G** | **Geography & Climate:** Environmental conditions and available materials. | `temp` (average temperature), `rain` (annual rainfall), `disrisk` (natural disaster risk), `mat` (local building materials), `construct_pref` (construction preference, e.g., modular, low-tech). |
| **SITE** | **Site Information:** Specific details of the construction site. | `size` (area in m²), `access` (road accessibility), `utilities` (on-site water/power), `topography`. |

## 3.2 Agent 2: Functional Programming Expert System Agent

This agent serves as the deterministic and verifiable "symbolic core" of the system. We designed a rule-based engine, implemented in Python, that receives the DQL and executes a series of rigorous, non-negotiable calculations to generate a precise, budget-compliant architectural functional program. Its logic is derived from established standards in public health and hospital planning, thereby shielding the core planning logic from the potential "hallucinations" of generative AI and ensuring the scientific rigor and reliability of design decisions.

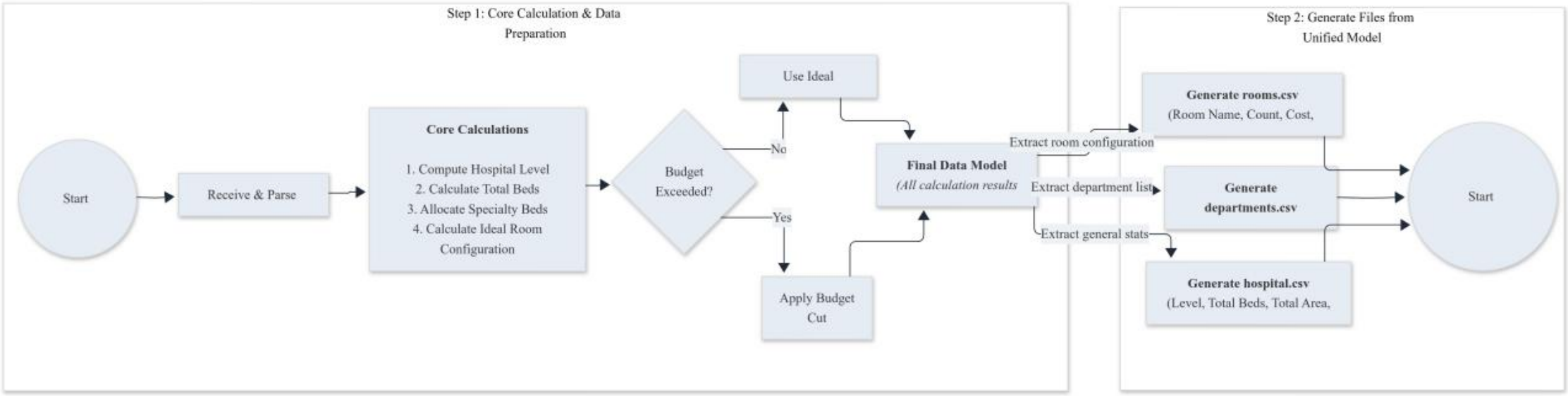

*Figure 7: Agent 2 Functional Planning Process*

1. **Population Projection & Hospital Level Determination**: The system's first task is to establish the scale of demand. It begins by projecting the service population for a given planning horizon (typically 10-20 years) to account for future growth.

$$P_{\text{projected}} = P_{\text{population}} \times \left(1 + \frac{P_{\text{growth_rate}}}{100}\right)^{\text{planning_years}}$$

Where:

$P_{\text{projected}}$ is the projected population.

$P_{\text{population}}$ is the current population from the DQL.

$P_{\text{growth_rate}}$ is the annual population growth rate.

Next, the system uses a multi-factor weighted scoring model to determine the appropriate level for the target hospital (e.g., Clinic, Primary, Secondary, Tertiary). This model moves beyond simple population thresholds to incorporate a nuanced understanding of the local context.

$$\text{Level Score} = w_p \cdot f(P_{\text{projected}}) + w_{\text{h}} \cdot f(H_{\text{complexity}}) + \sum_{i=1}^{n} w_i \cdot M_i$$

Where:

$w_p, w_{\text{h}}, w_i$ are weight coefficients.

$f(P_{\text{projected}})$ is a normalized score based on population size.

$f(H_{\text{complexity}})$ is a score derived from the complexity of the local disease profile (e.g., prevalence of diseases requiring specialized care).

$M_i$ are modifier factors drawn from the DQL, such as existing resources (E), infrastructure accessibility (I), economic capacity (X), and special risks like conflict (S).

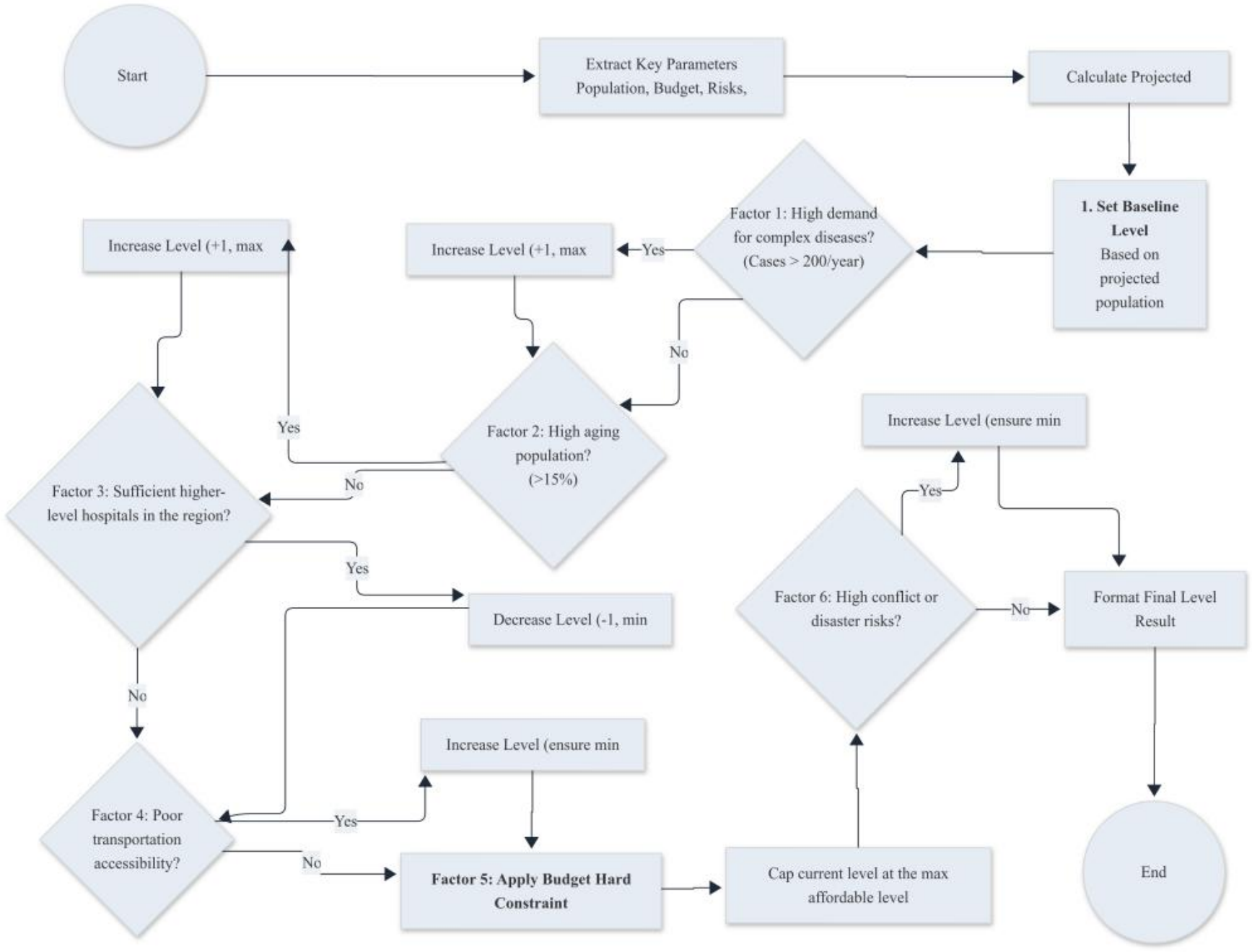

*Figure 8: Hospital Level Determination Logic*

2. **Net Bed Need Calculation**: The core of a hospital's scale is the calculation of bed demand. The expert system performs a multi-step calculation to determine the net bed gap that the new facility must address.

**Regional Theoretical Bed Need**: A baseline demand is established based on the projected population and a target bed rate (adjusted for the determined hospital level and regional benchmarks).

$$\text{Theoretical Total Beds} = \frac{P_{\text{projected}} \times \text{Target Bed Rate per 1000}}{1000}$$

**Existing Effective Bed Supply**: The system discounts the number of existing beds by a quality factor to account for poorly equipped or understaffed facilities, providing a more realistic measure of current capacity.

$$\text{Effective Existing Beds}=E_{\text{total_beds}}\times E_{\text{quality_factor}}$$

**Basic Net Bed Gap**: The fundamental shortfall in capacity.

$$\text{Net Bed Base}=\max(0,\text{Theoretical Total Beds}-\text{Effective Existing Beds})$$

**Additional Bed Demand**: The model incorporates modifier factors to add bed capacity for specific local needs identified in the DQL. This includes factors like maternal and child health needs ($M_{\text{health}}$), prevalence of specific infectious diseases, trauma rates from conflict or road traffic accidents ($S_{\text{conflict}}$,$S_{\text{vio}}$), and special care requirements.

**Target Total Bed Count**: The final bed count for the new facility is the sum of the base gap and all additional requirements, forming the basis for subsequent departmental planning.

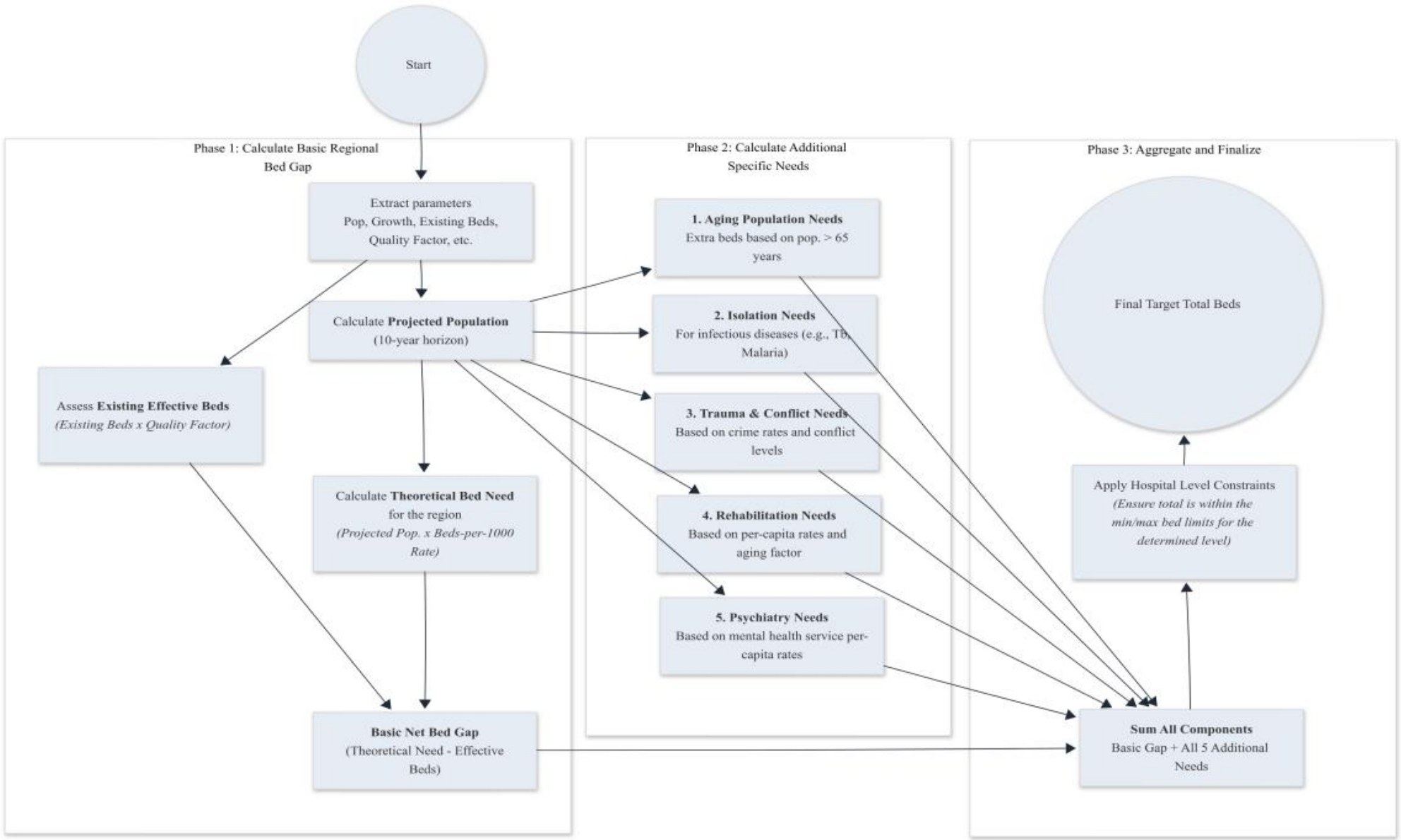

*Figure 9: Net Bed Need Calculation Process*

3. **Departmental Programming & Budget Optimization**: Once the total bed count is fixed, the system determines the required clinical and support departments based on the hospital level and the local disease profile (H). For each required department, it calculates the net need for specific functional spaces (beds, exam rooms, operating rooms, delivery rooms, etc.) using formulas derived from public health and hospital planning standards.

   For example, the net number of required operating rooms (ORs) is calculated using a standard surgical demand formula:

$$\text{Net ORs}=\max\left(0,\left\lceil\frac{\sum_s\ (\text{Beds}_s\times\text{Surgery Rate}_s)\times\text{Avg Duration}_s}{\text{Daily Op. Hours}\times\text{Annual Op. Days}\times\text{OR Utilization}}\right\rceil-E_{\text{ORs}}\right)$$

   Where:

   $s$ represents each surgical specialty.
   $\text{Beds}_s$ is the number of beds for that specialty.

Surgery $\text{Rate}_s$ is the annual number of surgeries per bed for that specialty.

Avg $\text{Duration}_s$ is the average duration of a surgery in hours.

OR Utilization is the target utilization rate of the ORs.

$E_{\text{ORs}}$ is the number of existing, functional ORs.

Finally, the system generates a complete room program listing the quantity and estimated area for every required room. It then calculates a total estimated construction cost based on regional cost-per-square-meter data, adjusted for local material availability (G) and labor costs (X). If this estimated cost exceeds the available budget ($X_{\text{budget}}$), the system applies an iterative optimization algorithm. It systematically trims or adjusts program elements according to a predefined hierarchy of clinical priorities (e.g., preserving critical care functions while reducing administrative space) until the design meets the budget constraint. This transparent, rule-based process ensures all final options are financially viable, transforming the "black box" of budgeting into an open ledger that all stakeholders can scrutinize and understand—a critical step in building **trust and reciprocity**.

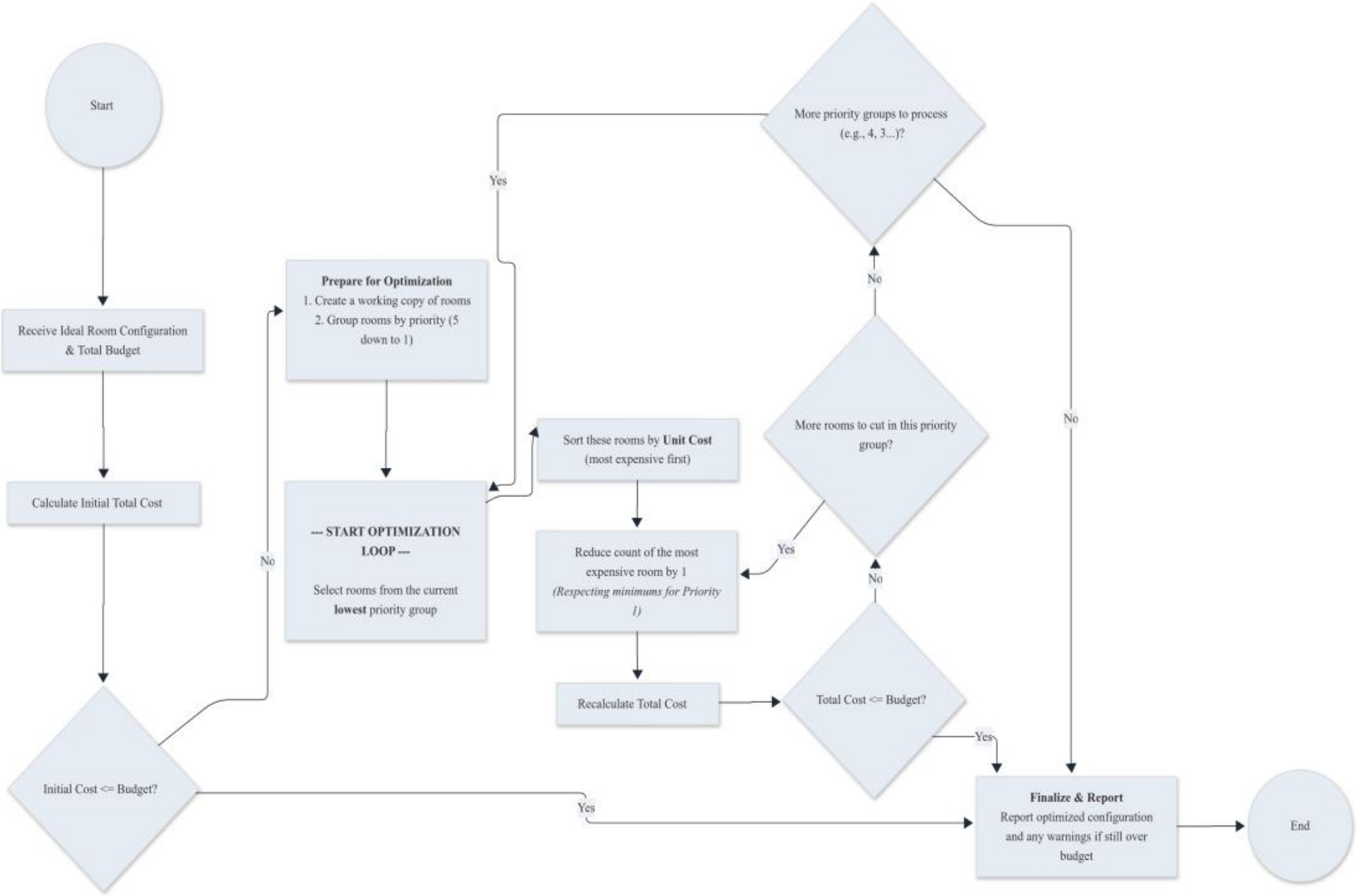

*Figure 10: Budget Optimization Algorithm*

For ease of reuse across countries/regions, the expert system employs a three-tiered abstraction model (Primary–Secondary–Tertiary, P/S/T) for functional programming. This model corresponds to the internationally common service delivery system tiered by function and referral pathways: primary care for continuity services as the entry point, secondary care for specialized/comprehensive enhancement, and tertiary care providing highly specialized and teaching/referral center functions. This abstraction aligns with the monitoring/evaluation frameworks of organizations like the WHO regarding service accessibility and referral systems, facilitating alignment with UHC-related indicators and regional benchmarks. [64]

### 3.3 Agent 3: Architectural Design & Layout Agent

This agent is responsible for translating abstract data into concrete spatial concepts, achieving a form of "constrained creativity." It consists of two sub-components: an LLM-driven Layout Planner and a procedural 3D Generation Engine. This agent not only generates aesthetically pleasing design options but, more importantly, ensures these options are technically feasible, functionally sound, and compatible with local construction conditions.

#### 3.3.1 Knowledge-Injected Generation (Axial Circulation Layout Planner)

The Layout Planner is invoked via a "knowledge-injected" prompt that provides the LLM with a comprehensive framework for hospital design. This prompt effectively acts as a condensed architectural treatise, providing the LLM with a comprehensive framework built on three pillars:

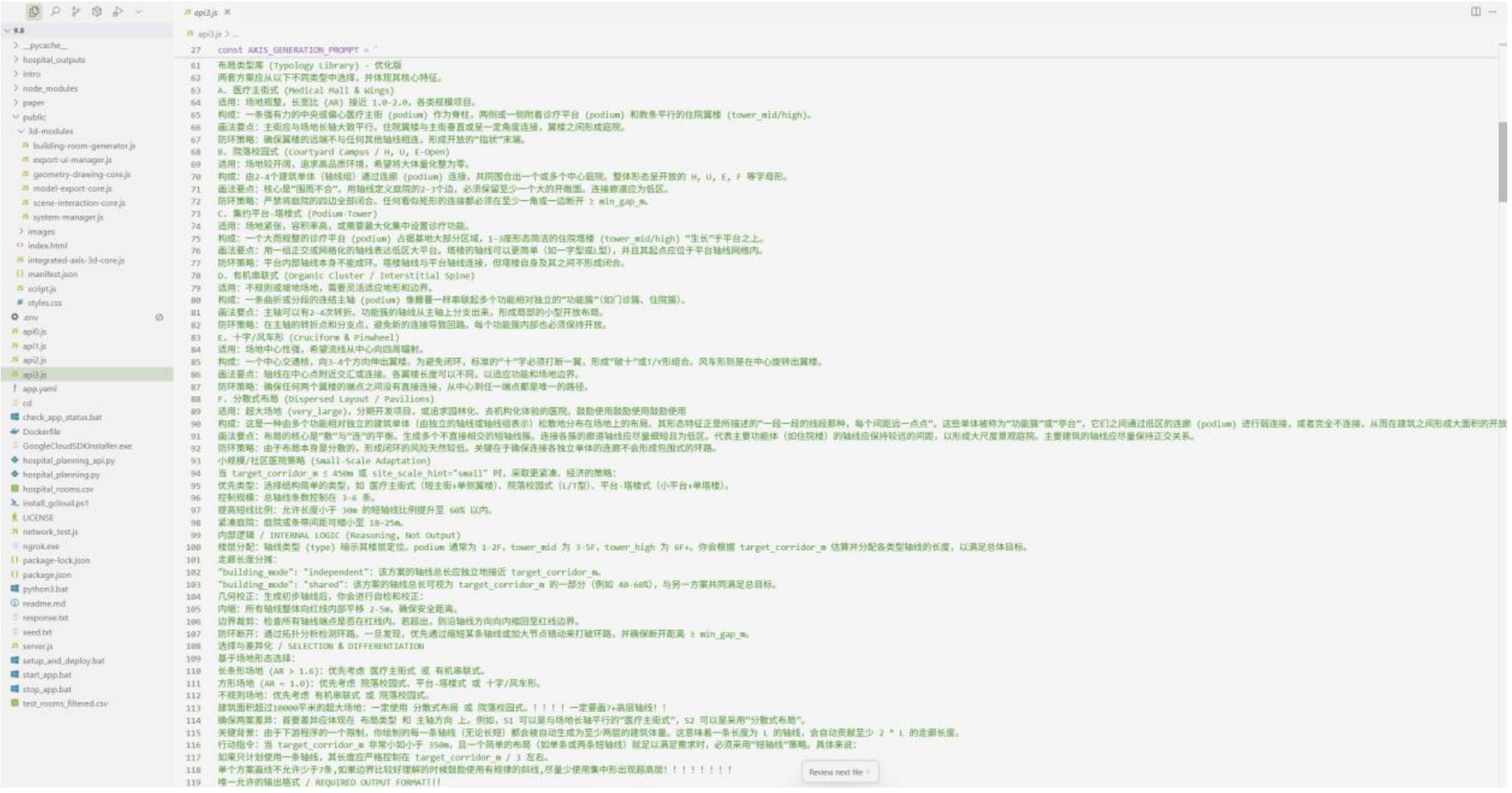

*Figure 11: Knowledge-Injected Generation*

**Typology Library:** To ground the LLM's generation in established architectural knowledge, the prompt provides a library of proven hospital layout archetypes. This constrains the vast solution space to professionally recognized patterns, ensuring the output is not just geometrically plausible but strategically robust. The model is instructed to select from these forms based on site conditions and project scale to encourage diversity in its proposals.

**Table 2: Hospital Layout Typology Library provided to the Layout Planner.** Layout typologies synthesized from architectural precedents [9] [65] [66].

| Typology Name | Description & Drawing Strategy | Key Features & Suitability |
|---|---|---|
| **Medical Main Street & Wings** | A strong central spine ("main street") axis ( `podium` ) connects various functional zones and inpatient wings. Wing axes are perpendicular or angled to the main spine. | Strong linear organization, clear circulation, facilitates phased expansion. Suitable for rectangular sites. |
| **Courtyard Campus** | Multiple building axes are arranged to "frame" but not fully enclose one or more open spaces, forming H, U, or E shapes. Connecting axes are low-rise ( `podium` ). | Promotes a healing environment, breaks down large massing, integrates landscape. Best for wide, open sites. |
| **Podium-Tower** | A dense grid of `podium` axes forms a large low-rise block for diagnostics/treatment, from which simple `tower_mid/high` axes "grow." | High density, spatially efficient, clear vertical separation of functions. Suitable for compact urban sites with high FAR requirements. |
| **Organic Aggregate** | A flexible, segmented, or curved primary axis ( `podium` ) links multiple smaller, functionally distinct clusters of axes. | Highly adaptable to irregular terrain, sloped sites, or complex site boundaries. |
| **Dispersed Pavilions** | Multiple disconnected short axes or small clusters of axes are distributed across the site, representing different pavilions, potentially linked by slender `podium` connecting axes. | De-institutionalized, "village" feel, maximizes landscape integration. Only suitable for very large sites or specialized facilities. |

**Core Design Principles:** The prompt encodes fundamental principles of hospital design, translating qualitative architectural goals into explicit instructions for the LLM. These include:

**Functional Zoning:** The model is instructed to differentiate axes by type (e.g., podium for low-rise public and diagnostic zones, tower_mid and tower_high for inpatient wings). This directly embeds the concept of separating public, clinical, and private zones into the generated data structure.

**Circulation & Flow:** The prompt emphasizes the need for clear circulation and the potential for separating public, staff, and service routes (the "three-corridor" concept). This directly influences the topology of the axis grid.

**Expandability & Phasing:** The agent is guided to favor open-ended, non-enclosed layouts (e.g., "finger" plans) to facilitate future expansion, a critical consideration for facilities in growing communities.

**Healing Environment:** The model is encouraged to arrange axes to form well-oriented, open courtyards that introduce natural light and green space, contributing to a therapeutic atmosphere.

**Solar Access & Spacing:** For multi-story elements, especially inpatient towers (tower_high), the prompt demands sufficient spacing to ensure adequate daylight and privacy, reflecting best practices in patient care.

**Constructability & Modularity:** The agent is explicitly prompted to favor geometries that are compatible with **modular and low-tech construction systems**. This includes a preference for regular orthogonal grids, rationalized structural spans that align with local material availability (e.g., timber beam lengths), and forms that are easily decomposable into repeatable components.

**Hard Geometric & Topological Constraints:** To ensure the validity and utility of the output, the prompt enforces a strict set of non-negotiable rules:

**Boundary Adherence:** All generated axes must fall entirely within the user-defined site boundary (outline). This is the highest-priority directive.

**Topological Integrity (Acyclicity):** The agent is strictly forbidden from creating closed loops in the axis grid. All schemes must be topologically equivalent to a tree or a forest, ensuring there is only one path between any two points in any connected component. This prevents the creation of dysfunctional enclosed courtyards and ensures clarity of wayfinding.

**Dual-Scheme Output:** The model must generate two distinct schemes, S1 and S2, which should differ significantly in their chosen typology, primary orientation, or topological structure. This provides the user with genuine alternatives.

The LLM is strictly constrained to generate its output in a precise JSON format that encodes geometric, hierarchical, and functional information:

```
{
 "schemes": [
   {
    "id": "S1",
    "building_mode": "shared",
    "axes": [
      {"start":[x_mm, y_mm], "end":[x_mm, y_mm], "type":"podium"},
      {"start":[x_mm, y_mm], "end":[x_mm, y_mm], "type":"tower_mid"},
      {"start":[x_mm, y_mm], "end":[x_mm, y_mm], "type":"tower_high"}
    ]
   },
   {
    "id": "S2",
    "building_mode": "independent",
    "axes": [
      {"start":[x_mm, y_mm], "end":[x_mm, y_mm], "type":"podium"},
      {"start":[x_mm, y_mm], "end":[x_mm, y_mm], "type":"tower_mid"}
    ]
   }
 ]
}
```

### 3.3.2 Procedural Geometry Engine

The 3D Generation Engine, built on Three.js, then receives this qualitative spatial data along with the quantitative room program from Agent 2. It uses a robust geometry kernel and a layered generative logic to transform the abstract datasets into concrete 3D spaces. This process is not just about geometric generation but about translating abstract design intent into a tangible architectural reality.

**Geometry Kernel: Robust Operations on a Discrete Integer Grid**

To circumvent the notorious instability of floating-point arithmetic in computational geometry (which can lead to catastrophic failures in Boolean operations), all core geometric calculations are performed on a discrete integer grid coordinate system $C \subset \mathbb{Z}^2$. A global scaling factor $S$ (e.g., 100) controls the translation between the real-world coordinate system $W \subset \mathbb{R}^2$ (in millimeters) and the integer grid $C$.

Mapping from world to grid coordinates: $f(p_w)=(\text{round}(p_w.x \cdot S),\text{round}(p_w.z \cdot S))$
Mapping from grid to world coordinates: $g(p_c)=(p_c.X/S,p_c.Y/S)$

Where $p_w=(x,z)$ is a point in world coordinates, and $p_c=(X,Y)$ is the corresponding integer grid point. This ensures that all data input to the polygon clipping library (Clipper.js) is integer-based, guaranteeing robust and error-free Boolean operations (union, difference, intersection).

**Contour Generation via Minkowski Sum Approximation**

The building footprint contour for each floor is generated by offsetting and unioning the set of AI-generated axes $A=\{A_1,A_2,...,A_n\}$ applicable to that floor. Geometrically, this operation is equivalent to the union of the Minkowski sum of each axis segment with a circular structuring element $D_r$, where $r$ is the room depth.

$$P_{\text{contour}}=\bigcup_{i=1}^{n}(A_i \oplus D_r)$$

Where $\oplus$ denotes the Minkowski sum operator. For computational efficiency on a discrete grid, we approximate this by generating a rectangle for each axis segment and then performing a union operation on the resulting set of rectangles. We use the Vatti clipping algorithm (as implemented in Clipper.js) for this. The output of this algorithm is a PolyTree data structure, which accurately describes complex polygonal shapes with internal holes (representing courtyards or atria)—a critical feature for building massing generation.

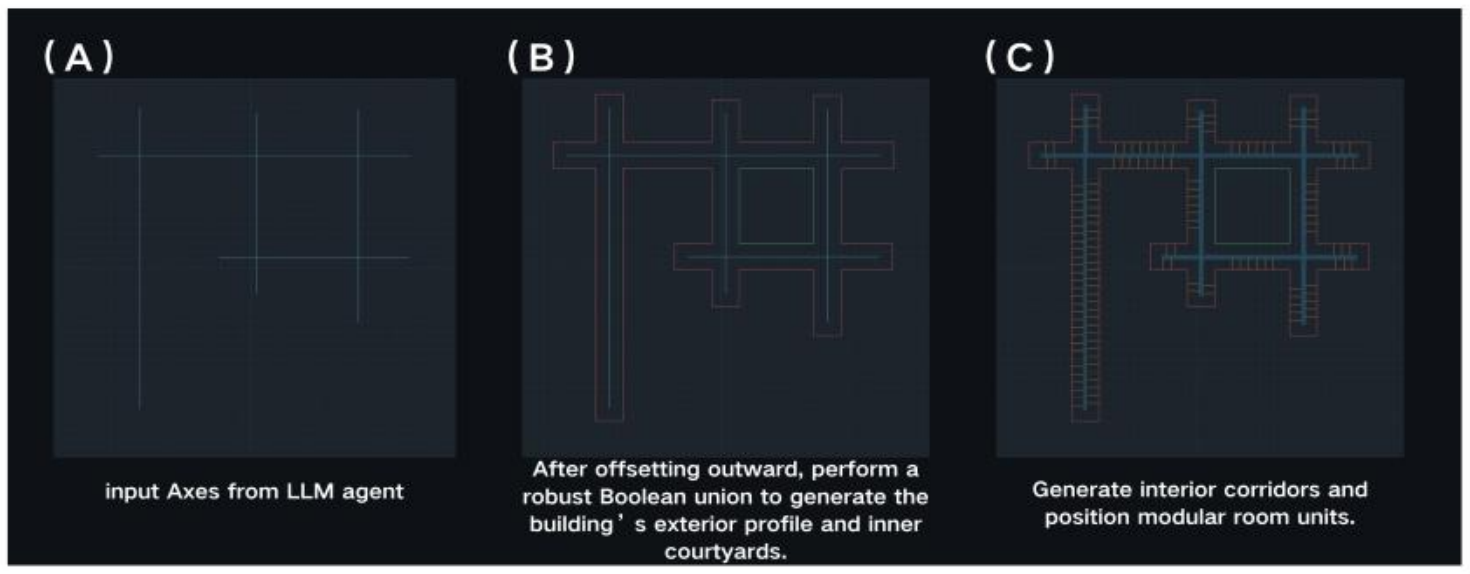

*Figure 12: Contour Generation Process*

**Core Spatial Layout Algorithm**

The layout algorithm forges the critical link between architectural form and functional program, ensuring the generated 3D model accurately matches the requirements calculated by the expert system. It operates via a three-stage process:

Intelligent Floor Planning: The system first determines the optimal number of floors required by dynamically calculating the capacity of each potential floor plate based on its available area. It continuously adds floors until the cumulative capacity significantly exceeds the total room requirement (e.g., reaching 130%), thus ensuring sufficient space while avoiding unnecessary vertical expansion. This stage establishes the building's fundamental vertical organization.

Waterfall Room Allocation: Once the floor plan is established, the system employs a "waterfall" allocation strategy. All rooms from the program are first sorted by clinical priority. Then, starting with the most critical functions, each room is sequentially assigned to the first available floor with sufficient capacity. This ensures that high-priority spaces are always placed first. After the initial allocation, any over-provisioned but still empty floors are removed from the final plan to optimize building efficiency.

Grid-Based Room Generation and Modular Synthesis: The final stage transforms the abstract, floor-assigned list of rooms into concrete 3D geometry. For each floor, the system first generates a structural grid based on the perimeter. The allocated rooms are then "packed" onto this grid, with each room being assigned a contiguous block of grid cells corresponding to its required area. The subsequent 3D synthesis is done in a uniquely modular fashion: instead of creating a single monolithic mesh for each room, the engine iterates through the grid cells assigned to a room and generates a separate geometric module for each cell. These modules are then combined to form the complete room. This approach ensures that all generated buildings are inherently modular, visually reinforcing their compatibility with prefabricated and modular construction systems and making the relationship between the underlying structural grid and the final spaces immediately legible.

**Algorithm 1: Core Spatial Layout Algorithm**

```
Input:
  RoomProgram: A list of rooms with quantities and areas
  ConceptualAxes: A set of axes with massing types (podium, tower_mid, tower_high)
  Parameters: Building parameters (room_depth, corridor_width, floor_height)

Output:
  BuildingModel: A complete 3D hospital model

// Stage 1: Intelligent Floor Planning
function CalculateOptimalFloorPlan(RoomProgram, ConceptualAxes, Parameters):
    totalRequiredRooms <- sum(room.quantity for room in RoomProgram)
    potentialFloors <- []
    cumulativeCapacity <- 0
    floorIndex <- 0

    while cumulativeCapacity < totalRequiredRooms * 1.3: // 130% overprovisioning for flexibility
        axesForFloor <- GetAxesForFloor(ConceptualAxes, floorIndex)
        if axesForFloor is empty and floorIndex > 0:
            break // No more massing on higher floors

        footprintArea <- CalculateFootprintArea(axesForFloor, Parameters.room_depth)
        corridorArea <- CalculateCorridorArea(axesForFloor, Parameters.corridor_width)
        usableArea <- footprintArea - corridorArea

        // Capacity based on a standard room module area
        floorCapacity <- floor(usableArea / STANDARD_ROOM_MODULE_AREA)

        potentialFloors.add({floorIndex, capacity: floorCapacity, axes: axesForFloor})
        cumulativeCapacity <- cumulativeCapacity + floorCapacity
        floorIndex <- floorIndex + 1

    return potentialFloors

// Stage 2: Waterfall Room Allocation
function AllocateRoomsToFloors(RoomProgram, FloorPlan):
    sortedRooms <- sort RoomProgram by clinical priority (descending)
    allocatedPlan <- copy of FloorPlan
    unallocatedRooms <- []


    for each room in sortedRooms:
        isAllocated <- false
        for each floor in allocatedPlan.potentialFloors:
            if floor.remainingCapacity >= room.area:
                floor.allocate(room)
                isAllocated <- true
                break

        if not isAllocated:
            unallocatedRooms.add(room)

    return allocatedPlan, unallocatedRooms

// Stage 3: Grid-Based Room Generation and Modular Synthesis
function Generate3DModel(AllocatedPlan, Parameters):
    model <- new 3DModel()

    for each floor in AllocatedPlan.potentialFloors:
        if floor.allocatedRooms is empty:
            continue

        // Generate structural grid for this floor
        grid <- GenerateStructuralGrid(floor.axes, Parameters)

        for each room in floor.allocatedRooms:
            // Pack room into grid cells
            gridCells <- PackRoomIntoGrid(grid, room.area, room.shape_preference)

            // Generate modular geometry for each cell
            for each cell in gridCells:
                module <- GenerateModuleGeometry(cell, room.type, Parameters)
                model.addModule(module)

    return model
```

Ultimately, the engine's primary role is not just to generate geometry, but to make the consequences of decisions legible. By providing a real-time, interactive environment, it transforms abstract data into a tangible spatial reality, thus creating a shared platform for designers, clients, and community members to negotiate the final design outcome. This visualization platform is the key instrument for achieving computational reciprocity, enabling all stakeholders to intuitively understand the consequences of design decisions.

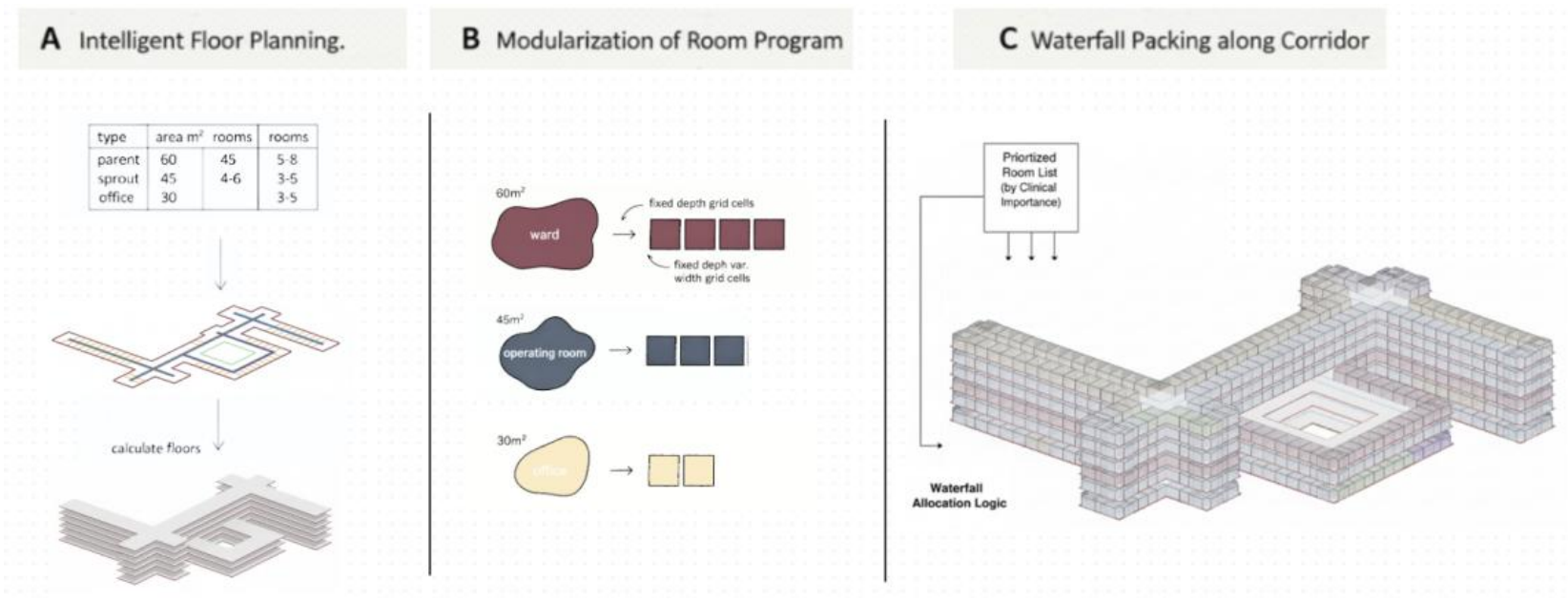

Figure 13: Procedural Geometry Engine Workflow.

## 4. Results and Analysis

To demonstrate MedBuild AI's end-to-end functionality and context-calibration capabilities, we present four diverse case studies derived from real-world scenarios. These scenarios showcase the platform's ability to tailor both functional programs and spatial outputs based on distinctly different local narratives.

### 4.1 Case Study 1: A Secondary Hospital for a Growing Township

This scenario represents a relatively well-resourced community poised for growth.

**Context:** A community of 50,000 with reliable infrastructure, active primary care, and a $5 million budget. The primary health challenges are a shift toward chronic diseases and a need for improved surgical and maternal care.

**Process:** The initial dialogue revealed competent local governance and clear clinical gaps. Agent 1 compiled this into a DQL reflecting a medium population, good infrastructure, and a specific budget: P:pop=50000,...|I:fac=0.8,...|X:budget=5,.... Agent 2, the expert system, calculated the need for a 172-bed secondary hospital and generated a detailed functional program, including 4 operating theaters and 8 ICU beds, costed to fit within the budget. Given the generous site, Agent 3 generated two spatially distinct concepts: S1, an efficient "Podium-Tower" scheme, and S2, a "Dispersed Campus" scheme that maximized green space for a healing environment.

**Outcome:** The system provided the community with two architecturally distinct, yet functionally identical and financially viable, options. This output serves not as a final solution, but as a data-driven starting point for **community-led negotiation**.

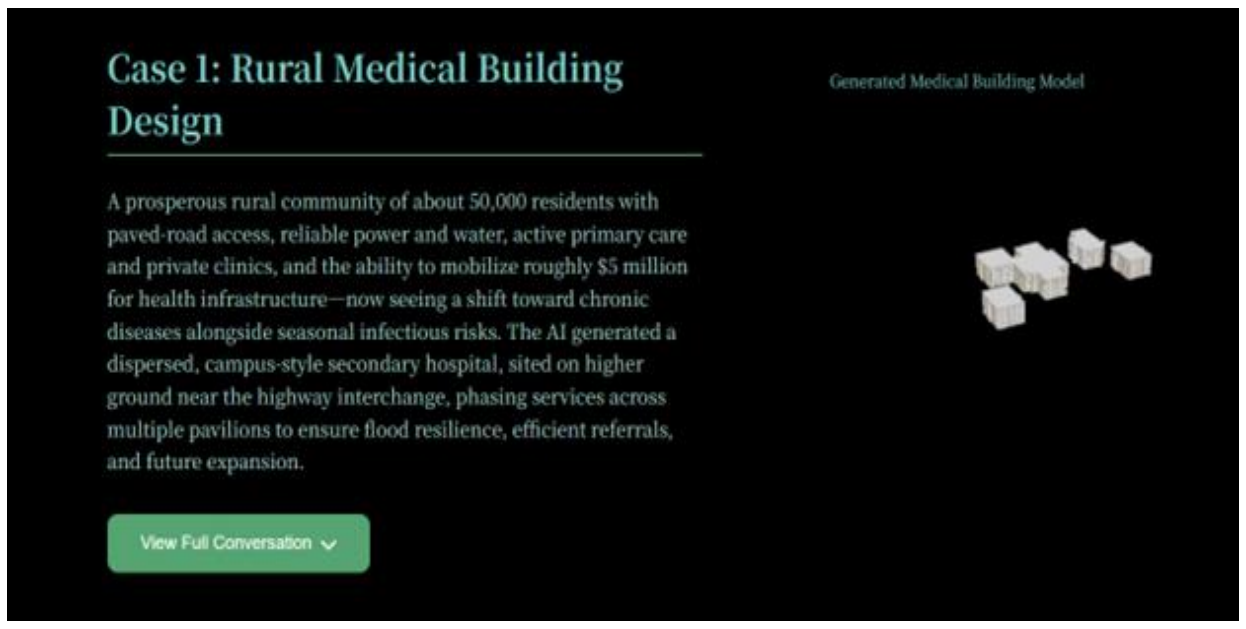

*Figure 14: Case Study 1 Results*

**Table 3: Abridged Room Program for Case Study 1**

| Department / Function | Quantity | Total Beds | Total Area (m²) |
|---|---|---|---|
| Outpatient Clinics | 23 | 0 | 350 |
| ORs & Delivery Rooms | 14 | 8 | 486 |
| Imaging Suites | 6 | 0 | 180 |
| Inpatient Beds (all depts.) | - | 172 | - |
| **Total** | **105** | **172** | **2,829** |

### 4.2 Cross-Scenario Comparative Analysis

To further validate the system's adaptability, we analyzed three additional scenarios with unique constraints. The results, summarized in Table 4, demonstrate the platform's ability to appropriately scale its outputs across a wide spectrum of needs and resources.

**Table 4: Comparative Analysis of Generated Scenarios**

| Case | Context & Key Challenges | Population | Budget | Hospital Level | Beds | Area (m²) | Spatial Concept |
|---|---|---|---|---|---|---|---|
| **1: Township** | Affluent, good infrastructure. Shift to chronic disease. | 50k | $5M | Secondary | 172 | 2,829 | Dispersed Campus |
| **2: Major Metropolis** | Dense informal settlement, unreliable utilities, trauma. | 600k | $50M | Tertiary | 1,500 | 26,294 | High-rise Podium-Tower |
| **3: Future New City** | High investment, rapid growth, world-class standards. | 3M | $800M | Tertiary | 1,500 | 40,348 | Mega-Campus |
| **4: Coastal Village** | Remote, poor infrastructure, climate risk (cyclones). | 5k | $0.8M | Primary+ | 10 | 544 | Compact, modular building |

The analysis shows that MedBuild AI successfully tailored both the scale of the functional program and the nature of the architectural form to the initial narrative. For the major metropolis (Case 2), it generated a high-density, 1500-bed tertiary hospital. For the ambitious new city (Case 3), it generated a massive teaching hospital. Crucially, for the vulnerable coastal village (Case 4), it correctly scaled down to a small, resilient 10-bed primary health center and proposed a compact, elevated structure well-suited for modular, low-tech construction and resilient to flooding. This demonstrates the system's ability to generate contextually appropriate solutions through a logical, context-driven form-generation process.

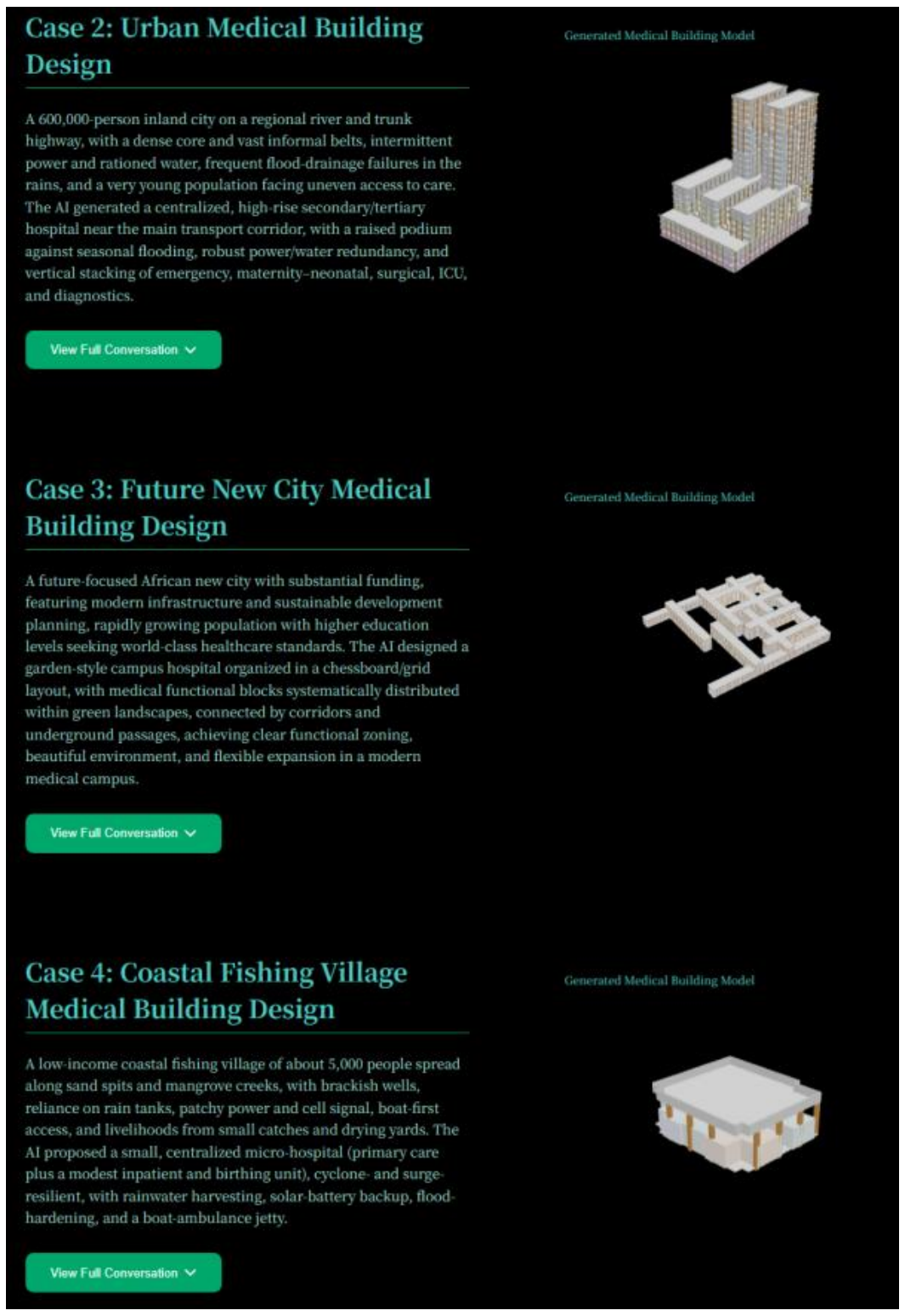

*Figure 15: Cases 2-4 Models*

## 5. Discussion

### 5.1 Theoretical Contribution & Technical Paradigm Comparison

Using MedBuild AI as a case study, this paper reflects on the role of computation in architectural design, especially in complex contexts of resource constraints and power imbalances. We attempt to construct a hybrid intelligence framework that converges the strengths of various single-path methods and provides an operational computational pathway for classic design theories like Function-Behavior-Structure (FBS) [19]. In the field of medical architecture design, existing computational methods are diverse but have their own shortcomings [16]: operations research-driven methods (e.g., QAP, MIP) excel at finding optimal solutions for quantitative objectives but rely on highly structured inputs (e.g., REL-Charts) and are insufficient in characterizing the ambiguity of cultural and social needs [12],[11]; rule-based generative systems (e.g., Shape Grammars) are interpretable and can formalize expert knowledge, but their rigid rules limit flexibility and creativity, making it difficult to incorporate unexpected local demands [28]; heuristic methods (e.g., evolutionary algorithms) can expand the search space to optimize adjacency relationships, but their process is stochastic, and optimality and full interpretability are hard to guarantee. Recent neuro-symbolic approaches have attempted to use the semantic understanding of LLMs to generate constraints for symbolic solvers, aiming to bridge the gap between quantitative and qualitative aspects [58]; progress in geometric theorem proving also suggests their feasibility in rigorous logical tasks [59]. Based on this line of thought, MedBuild AI combines the semantic representation of LLMs with the deterministic logic of an expert system, striving for a balance between creativity and verifiability, and demonstrating the potential of neuro-symbolic AI in handling complex spatial and logical problems [42],[43]. At the same time, this research focuses on a clear socio-technical orientation: supporting the negotiation process through a conversational front-end and interactive, clear outputs, with a focus on reshaping agency [13].

To more clearly position MedBuild AI, the following table compares the capabilities of recent computational design frameworks under different methodologies for key tasks in healthcare facility planning:

## Table 5: Functional Comparison of Computational Design Frameworks for Medical Architecture

| | Mixed-Integer Programming / Linear Programming (MILP / MIP) | Quadratic Assignment Problem (QAP) | Shape Grammars (Shape Grammar) | Industrial rule systems / scoring models | Genetic Algorithms (GA) | Multi-objective evolutionary algorithms (NSGA-II) | GAN generative models | Generative optimization / resilience design | Neuro-symbolic hybrid intelligence (LLM/Neuro-Symbolic) | IFC conversion models | Space syntax / metric studies | Hybrid intelligence |
|---|---|---|---|---|---|---|---|---|---|---|---|---|
| References | [11] | [22] | [27] | [26] | [23] | [67] | [68] | [69] | [43] | [70] | [40] | This paper |
| Publication year | 2022 | 2024 | 2020 | 2016 | 2024 | 2021 | 2021 | 2024 | 2025 | 2025 | 2020 | 2025 |
| Input form | Structured parameters; REL chart or REL diagram; geometric | Department/room set; adjacency frequency matrix; geometric grid | Rule library; initial graph; component dictionary | Score sheet; weighted rules; parameter list | Parametric geometry; objective function; constraint | Parametric model; multi-objective weights | Image/raster; few condition labels | Parametric layout; pedestrian/emergency parameters | Natural language plus structured tables; knowledge base | IFC/JSON; model conversion configuration | Floor plan/graph theory; spatial network | Natural language plus site redline |
| Intent understanding (handling fuzzy/qualitative input) | Weak; requires precise quantitative inputs | Weak; discrete optimization struggles with ambiguity | Medium; semantic labels encodable | Weak; subjective scoring hard to interpret | Weak; quantized to numbers | Weak; needs quantified objectives | Weak; poor semantic control | Weak; relies on numeric settings | Strong; LLM negotiative understanding | Weak; not an intent system | Weak; analysis not understanding | Strong; dialogic negotiation; context memory |
| Design brief generation (quantified room program) | Weak; relies on external inputs | Weak; room counts predetermined | Medium; generated via rules | Weak; mostly evaluative not generative | Weak; does not generate explicit list | Weak; not responsible for generation | Weak; hard to directly output a list | Medium; adjusted with simulation | Strong; rule core precise computation | Weak; not responsible for generation | Weak; no program list output | Strong; expert-system hard-constraint computation |
| Floorplan / circulation planning | Strong; global optimization of walking and logistics | Strong; minimizes walking and logistics | Medium; depends on rule refinement | Medium; scoring guides scheme selection | Medium; approximated via fitness | Medium; optimizes along metrics | Weak; lacks explicit circulation | Strong; emergency-oriented circulation | Medium; concept layout plus rule constraints | Weak; not a planning module | Medium; evaluation not generation | Medium; concept layout plus rule correction |
| Multi-storey generation | Medium; stackable solving | Weak; mostly single-layer grids | Medium; rules extendable | Weak; lacks general approach | Medium; iterative extension | Weak; mostly single-level cases | Weak; mostly single-layer | Medium; extendable to multi-storey | Strong; modular synthesis | Weak; not a generative module | Weak; analysis mostly single-layer | Strong; modular floor synthesis |
| Constraint handling | Hard constraints; solver strictly satisfies | Hard constraints; penalty terms assist | Hard rules plus sequencing; composable | Soft constraints; weighted scoring | Soft constraints mainly; penalty terms | Soft constraints plus dominance relations | Weak; corrected in post-processing | Hybrid soft/hard; simulation calibration | Hybrid; rules hard constraints plus LLM soft guidance | Medium; structured mapping validation | Weak; lacks generative constraints | Hybrid; rules hard constraints plus LLM soft guidance |
| Performance metrics / optimization | Single; distance or cost dominated | Single or bi-objective; distance dominated | Weak; mostly constraint satisfaction | Multi-metric; weights subjective | Multi-objective extensible | Strong; Pareto front | Weak; aesthetic/statistical similarity | Strong; resilience and efficiency balanced | Medium; can incorporate multiple metrics | Weak; no optimization objective | Strong; integration and accessibility metrics | Medium; can attach multi-metric dashboard |
| Process efficiency | Medium; time-consuming at scale | Medium; solved by approximations | High; rule derivation fast | High; simple computation | Low; iterative cost high | Low; compute cost high | High; inference fast training slow | Medium; simulation time-consuming | Medium; inference plus rule computation | High; efficient batch conversion | High; metrics compute quickly | Medium; two-stage inference plus validation |
| Accessibility and visibility | Medium; can include but costly | Weak; rarely models visibility | Medium; checks can be embedded | Medium; metrics can be added | Medium; spatial analysis can be added | Medium; combined analyses possible | Weak; needs external analysis | Medium; visibility can be added | Medium; plug-in spatial analysis | Medium; relies on external analysis | Strong; core metrics | Medium; can hook visibility/path analysis |
| Cost and benefit | Medium; seeks optimum but modeling costly | Medium; uses operational distance as proxy | Medium; high upfront rule cost | Medium; depends on weight settings | Medium; high computational cost | Medium; needs compute resources | Medium; large data requirement | Medium; multi-objective trade-offs | Medium; depends on LLM/rule engineering | High; standardization enables reuse | Medium; needs pairing with a generator | Medium; built-in budget optimization |
| Digital twin and decision support | Medium; simulation APIs easy to plug | Weak; limited simulation integration | Medium; can integrate evaluators | Medium; panel-based decisions possible | Medium; can link with simulations | Medium; cutting-edge visual decision-making | Weak; lacks interpretable metrics | Strong; simulation-driven decisions | Strong; human–AI co-creation decisions | Medium; data base for the twin | Medium; evaluation supports decisions | Strong; negotiation and version comparison |
| Generative diversity | Low; tends to single optimum | Low; similar solution sets | Medium; determined by rule diversity | Weak; more like a filter | High; rich solution set | High; non-dominated set | High; high style variation | Medium; depends on optimizer | High; prompts guide diverse outputs | None; not generative | None; not generative | High; multi-scheme dialogic generation |
| Output interpretability | High; mathematical optimum traceable | High; objective functions clear | High; rules transparent | Medium; but lacks peer review | Medium; stochastic search | Medium; needs visualization | Low; black box | Medium; metrics transparent model complex | Medium; rules transparent LLM opaque | High; standards readable | High; metrics explicit | Medium; rules transparent LLM opaque |
| Output format generality | Medium; exportable but needs further development | Weak; needs post-processing | Medium; needs component mapping | Weak; no standard export | Medium; components exportable | Medium | Weak; needs vectorization/parsing | Medium; exportable | Medium; needs mapping to generate IFC | Strong; native IFC | Weak; needs save-as | Medium; mappable export to common 3D formats |
| Compliance (three zones and two passages) | Medium; hard constraints encodable | Weak; detailed industry rules hard to encode | Medium; expressible via rules | Medium; clauses can be added | Medium; expressed via constraints | Medium | Weak; cannot directly enforce constraints | Medium; passage rules encodable | Medium; can encapsulate rules | Weak; needs external validation | Weak; not industry rules | Medium; expressed via rule encapsulation |
| Geometric generalization ability | Medium; depends on parametric boundary | Weak; limited by grid | Medium; limited by rule coverage | Weak; not directly modeling geometry | Medium; depends on encoding | Medium | Medium; strong visual generalization weak geometry | Medium; depends on parametric modeling | Strong; templates plus generation | Medium; independent of generator | Medium; depends on input plan | Strong; axes plus grid synthesis |
| Cross-site and complexity adaptation | Medium; costly to rebuild model | Medium; transferable by changing parameters | Medium; needs new rules to adapt | Medium; adjustable weights | Medium; parameters transferable | Medium | Medium; domain-dependent data | Medium; parameters transferable | Strong; contextual adaptation | High; standards equal adaptation | High; universal metrics | Strong; context-adaptive and templated |
| Reproducibility and dependencies | High; depends on commercial/open-source solvers | High; algorithms public | High; rules reusable | Low; report-level materials | Medium; strong randomness | Medium | Medium; training depends on GPU | Medium; depends on simulation platform | Medium; depends on model version | High; toolchain stable | High; mature methods | Medium; depends on LLM version and rule base |
| Example of generated floor plans | 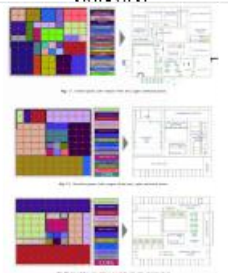 | 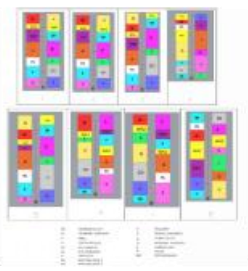 | 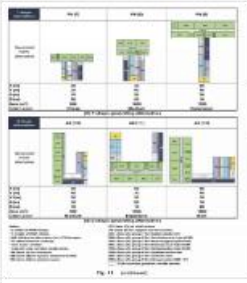 | 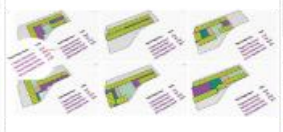 | 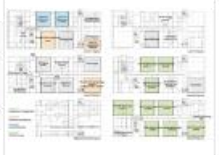 | 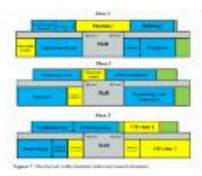 | 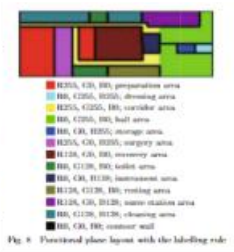 | 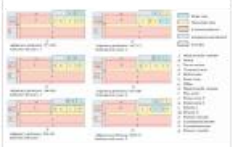 | Yes，but no images were found. | No | No | 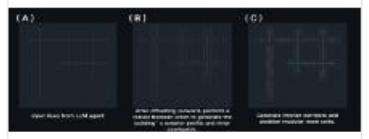 |
| Example of generated 3D models | 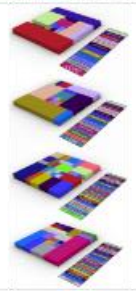 | NO | NO | NO | 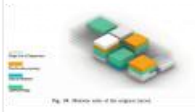 | NO | NO | NO | Yes，but no images were found. | No | No | 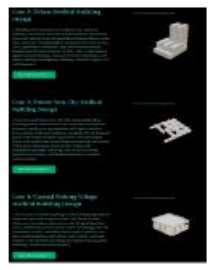 |

As seen from the table, MedBuild AI aims to integrate the advantages of different methods: it possesses the intent understanding capability of neuro-symbolic frameworks and deepens it from passive text parsing to active conversational negotiation; it has the program generation capability of a deterministic expert system, ensuring the rigor and verifiability of the solutions; at the same time, it makes multi-objective optimization its core social function—presenting the conflicting needs of different stakeholders (such as cost, cultural appropriateness, clinical efficiency) in a visual 3D form, thereby transforming the design process itself into a transparent negotiation platform.

### 5.2 Reshaping Agency and the Design Workflow

The platform fundamentally reshapes the pre-design workflow, transforming it from a linear, opaque, expert-driven process into a rapid, transparent, and collaborative cycle. By making the causal link between input (local knowledge) and output (spatial options) immediately visible, it shifts the architect's role from an isolated "solution provider" to a "facilitator of dialogue" and "negotiator of trade-offs," which resonates with critical perspectives on the nature of participation in design [63], [38]. The framework encourages a deeper level of human-AI interaction, where the system augments rather than replaces decision-making [61]. This process actively reshapes agency, empowering local, non-expert stakeholders to engage in foundational design decisions on an evidence-based platform, thus turning a design tool into a genuine platform for negotiation [14], and also provides a potential path for incorporating more diverse inclusive design considerations in the future (e.g., assistive design for people with disabilities) [62].

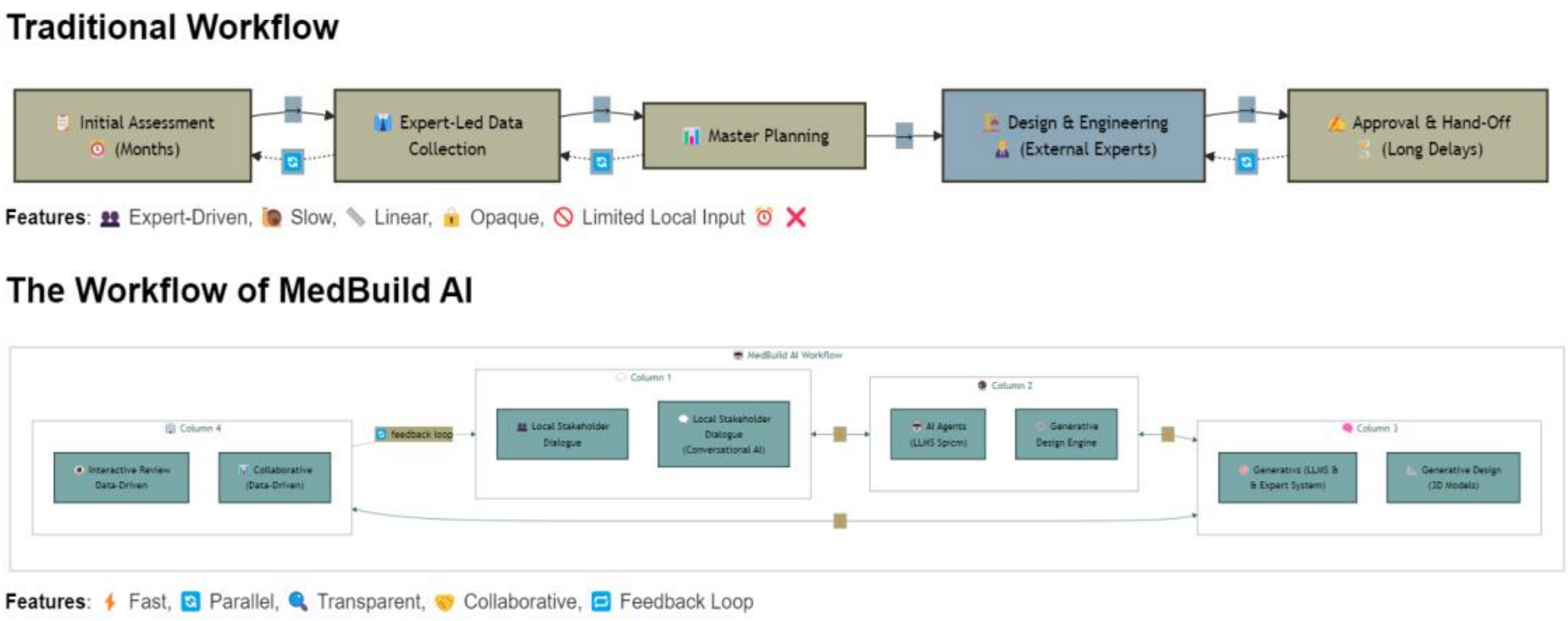

*Figure 16: Traditional vs MedBuild AI Workflow*

### *5.3 Limitations and Future Directions*

Despite its theoretical novelty, MedBuild AI's technical implementation still has significant limitations. The PHCMEISXG framework, though comprehensive, cannot fully capture the complexity and dynamism of culture; the current axis generation system, while creative, has limitations in handling complex terrain and special functional requirements. More importantly, the system lacks the ability to simultaneously optimize multiple conflicting objectives, which limits its application in complex decision-making scenarios. Furthermore, the system's deployment faces significant socio-political challenges: a standardized workflow might erode the uniqueness of local cultures, which requires us to pay more attention to cultural sensitivity and adaptability in future development.

Future development will focus on enhancing the system's global adaptability and multi-objective optimization capabilities. First, by establishing a dynamic knowledge base containing healthcare standards and building codes from different countries to support the system's deployment in various cultural contexts, and developing a more granular system of cultural parameters capable of capturing complex cultural factors such as religious beliefs, social structures, and traditional medicine. Second, to address the limitations of the current spatial generation system, the focus will be on optimizing the axial circulation layout design system, developing graph theory-based axis network optimization algorithms to support more complex spatial topological relationships, and establishing a richer library of hospital building typologies, including specialized hospitals, rehabilitation centers, and community health centers.

Finally, referencing the Space Plan Generator method by Das et al. [26], a multi-agent multi-objective optimization framework will be developed. This system will simultaneously optimize multiple objectives such as cost, functional efficiency, environmental performance, and cultural appropriateness, with different agents representing the needs of different stakeholders, reaching a balanced solution through negotiation. This can be combined with machine learning methods for building performance (e.g., energy) modeling [36]. Drawing on the latest graph neural network technology, GNNs will be used to learn functional relationships and spatial adjacency patterns between rooms from a large number of hospital floor plans, supporting layout optimization based on user feedback [35]. This will enable the system to better understand complex spatial relationships, generate more rational layout schemes, and provide decision-makers with an intuitive display of multi-objective trade-offs, transforming layout design into an explicit negotiation of priorities.

## 6. Conclusion

Future development will focus on deepening the system's intelligence, enhancing its professional integration capabilities, and strengthening its ability to support more granular negotiation. A key direction is to improve transparency and trust by integrating a formal knowledge graph of healthcare standards—akin to a digital, queryable building code platform [60]—enabling the system to cite the specific regulations underpinning its calculations using Retrieval-Augmented Generation (RAG) techniques. To deepen spatial intelligence, we plan to introduce Graph Neural Networks (GNNs) to analyze the room program as a relational graph, optimizing functional adjacencies and presenting a Pareto front of solutions to make the negotiation of priorities more explicit [35]. Finally, to bridge the gap with professional practice, a critical next step is to upgrade the output to a preliminary Building Information Model (BIM) in the open IFC format [55], thereby creating a seamless connection to downstream detailed design and to healthcare building evaluation and optimization systems based on IFC-to-HCM conversion [41].

In conclusion, MedBuild AI demonstrates a novel hybrid intelligence framework not just for design generation, but for design negotiation. By orchestrating a reciprocal workflow between a multi-agent LLM pipeline, a deterministic expert system, and a procedural geometry engine, the system offers a robust and responsible approach to reshaping complex planning dynamics in resource-constrained contexts. The proposed "dialogue-compile-calculate-generate-synthesize" workflow offers a new paradigm for fostering computational reciprocity in the pre-design phase of other complex building types [15]. This work marks a significant step in the application of AI in architecture, moving from discrete tools to integrated systems that support critical reflection and shared decision-making. Most importantly, it offers a technologically empowered response to the immense challenges of global health, not by imposing solutions, but by facilitating a more balanced, reciprocal, and computationally-empowered negotiation of our built future.

## 7. Acknowledgements

The authors used generative-AI tools (OpenAI ChatGPT; Google Gemini) to assist with idea exploration, drafting, code comments, and layout-option summaries. All AI-assisted content was verified and edited by the authors; references were checked against primary sources; AI tools were not listed as authors. We acknowledge inspiration from the work of Moritz Rietschel and Kyle Steinfeld on reasoning-model-driven computational design; early prototypes benefited from their tooling in accelerating elements of the 3D generation pipeline. We thank Prof. Feng Yuan (Tongji University) for organizing DigitalFUTURES, where formative discussions took place; Prof. Yi Qi (Shenzhen University), who led another DigitalFUTURES studio and provided helpful suggestions; Prof. Kyle Steinfeld (University of California, Berkeley) for his DigitalFUTURES workshop lecture on generative design; and Ramon Weber (University of California, Berkeley) for a workshop lecture on generative computational design with LLM agents—both of which informed our approach. We are also grateful to the management partners of Zhonggang Hospital (DRC) for advice on the project's medical-context background, and to Jiaqi Tan (Tianjin University), Xiang Li (The Hong Kong Polytechnic University), and Khadijia Aliyu (BUCEA) for insightful comments on early drafts. Any errors remain our own.